\documentclass[fleqn,usenatbib]{mnras}

\usepackage{newtxtext,newtxmath}
\usepackage[T1]{fontenc}

\DeclareRobustCommand{\VAN}[3]{#2}
\let\VANthebibliography\thebibliography
\def\thebibliography{\DeclareRobustCommand{\VAN}[3]{##3}\VANthebibliography}

\usepackage{graphicx}	% Including figure files
\usepackage{amsmath}	% Advanced maths commands
\title[A moderate spin for MAXI J1348-630]{A moderate spin for the black hole in X-ray binary MAXI J1348-630 revealed by Insight-HXMT}

\author[Wu et al.]{
Hanji Wu,$^{1,2}$
Wei Wang,$^{1,2}$\thanks{E-mail: wangwei2017@whu.edu.cn}
Na Sai,$^{1,2}$
Haifan Zhu,$^{1,2}$
and Jiashi Chen$^{1,2}$
\\
$^{1}$Department of Astronomy, School of Physics and Technology, Wuhan University, Wuhan 430072, China\\
$^{2}$WHU-NAOC Joint Center for Astronomy, Wuhan University, Wuhan 430072, China\\
}

\date{Accepted XXX. Received YYY; in original form ZZZ}

\pubyear{2023}

\begin{document}
\label{firstpage}
\pagerange{\pageref{firstpage}--\pageref{lastpage}}
\maketitle

\begin{abstract}
MAXI J1348-630 is a low-mass X-ray black hole binary located in the Galaxy and  undergone the X-ray outburst in 2019. We analyzed the observation data in very soft state during the outburst between MJD 58588 and MJD 58596 based on the Insight-HXMT observations from 2 -- 20 keV via the continuum fitting method to measure the spin of the stellar-mass black hole in MAXI J1348-630. The inner disk temperature and the apparent inner disk radius were found to be $0.47\pm 0.01 \rm keV$ and $5.33\pm 0.10 \ R_{g}$ from the observation data modeled by the multicolor disc blackbody model. Assuming the distance of the source $D\sim 3.4 \rm kpc$, the mass of the black hole $M\sim 11 \ M_{\odot}$, and the inclination of the system $i\sim 29.2^{\circ}$, the spin is determined to be $a_{\star}=0.41\pm 0.03$ for fixing hardening factor at 1.6 and $n_{H}=8.6\times 10^{21} \rm cm^{-2}$. Besides, considering the uncertainty of the parameters $D, M, i$ of this system, with the Monte Carlo analysis, we still confirm the moderate spin of the black hole as $a_{\star}=0.42^{+0.13}_{-0.50}$. Some spectral parameters (e.g., column density and hardening factor) which could affect the measurements of the BH spin are also briefly discussed.
\end{abstract}

\begin{keywords}
X-rays: binaries -- accretion discs -- X-rays : individual (MAXI J1348-630)
\end{keywords}

%%%%%%%%%%%%%%%%%%%%%%%%%%%%%%%%%%%%%%%%%%%%%%%%%%

%%%%%%%%%%%%%%%%% BODY OF PAPER %%%%%%%%%%%%%%%%%%

\section{Introduction}

The black hole (BH) spin will reveal their formation and evolution and affect fruitful astrophysical phenomena. The spin would be helpful for understanding the relativistic jets launched by Blandford–Znajek (BZ) mechanism \citep{blandford1977electromagnetic} or Blandford–Payne (BP) mechanism \citep{blandford1982hydromagnetic}, and the ratio between the energy of accretion matter and electromagnetic radiation could vary by almost one order of magnitude, which depends on how rapidly the black hole rotates \citep{bardeen1972rotating}. For exploring the black hole spin $a$, the dimensionless spin parameter $a_{\star}$ is used to characterize the spin in different mass black holes, defined as $a_{\star}\equiv\frac{a}{M}=\frac{cJ}{GM^{2}}$ \citep{kerr1963gravitational,misner1973gravitation}, where the $M$ and $J$ represent the mass and the angular momentum of the black hole, the $c$ for light speed, $G$ for gravitational constant. In an accreting black hole, the spin $a_{\star}$ will shape the temperature of the inner disk \citep{zhang1997black,mcclintock2013black}, due to the spin-dependent innermost stable circular orbit (ISCO) \citep{bardeen1972rotating}, so that the more binding energy from the accreting matter heats the inner disk reaching a higher temperature. In an accreting black hole X-ray binary \citep{remillard2006x}, the outburst usually obeys the "Q"-shape feature evolving from the low hard state (LHS) though the intermediate state (IS) to the high soft state (HSS), then back to LHS. The theoretical simulations \citep{shafee2008three,penna2010simulations} suggest that the inner disc radius $(R_{in})$ would extend to the ISCO in the high soft state, the inner disc radius $R_{in}$ would not change \citep{steiner2010constant,kulkarni2011measuring}. 

At present, there are two main methods to measure the BH spin, namely the Fe K$\alpha$ emission line method which reflects the characteristic of reflected fluorescence in the
disk \citep{fabian1989x,dabrowski1997profile,reynolds2003fluorescent}, and the continuum-fitting (CF) model based on the Novikov–Thorne thin disc model \citep{novikov1973astrophysics,zhang1997black,li2005multitemperature}. The basic assumption of the CF method is the accretion disc extending into the innermost stable circular orbit (ISCO), and in order to derive a credible spin by the CF method, the luminosity must also be limited to below 30\% of the Eddington limit to ensure the application approximation of thin disk assumption.  The CF method has measured the spin of many stellar-mass BHs \citep{guan2021physical,zhao2021estimating,shafee2005estimating,steiner2011spin,mcclintock2006spin,zhao2020confirming}. In this paper, we will use the CF method to constrain the black hole spin of MAXI J1348-630.

%So the intrinsic disc luminosity is proportional to the fourth power of the inner disk temperature \citep{mcclintock2006spin}.

The Gas Slit Camera (GSC) onboard Monitor of All-sky X-ray Image discovered MAXI J1348-630 on Jan 26 2019 \citep{matsuoka2009maxi,yatabe2019maxi}, which is classified as a black hole X-ray binary (BHXRB) through the mass estimate, light curve evolution and spectral features \citep{tominaga2020discovery,belloni2020time,carotenuto2019meerkat,denisenko2019optical,jana2019preliminary,kennea2019maxi,russell2019atca,sanna2019nicer,yatabe2019maxi}. The MAXI J1348-630 obeyed the "Q"-shape feature during the 2019 outbursts in the Swift/XRT \citep{tominaga2020discovery} data and the NICER data \citep{zhang2020nicer}.  A giant dust scattering ring around this source is discovered in SRG/eROSITA data, and with the joint data from XMM-Newton, MAXI, and Gaia, \cite{lamer2021giant} estimated the distance of MAXI J1348-630 at 3.39 $\pm$ 0.34 kpc and the black hole mass of 11 $\pm 2 M_{\odot}$. From the spectral evolution during the 2019 outburst using the Swift and MAXI detectors, the two-component advective flow model estimated the black hole mass to be $9.1^{+1.6}_{-1.2}M_{\odot}$ \citep{jana2020accretion}. The Australian Square Kilometre Array Pathﬁnder (ASKAP) and MeerKAT also provided the observation of MAXI J1348-630 and showed that the H \uppercase\expandafter{\romannumeral1} absorption spectra of this source suggested the distance as $2.2^{+0.5}_{-0.6}$ kpc \citep{chauhan2021measuring} and \cite{carotenuto2022modelling} modeled the physical properties of jets from MAXI J1348-630. 

The inclination angle of the binary system is also an important parameter for black hole research. \cite{carotenuto2022modelling} obtained the inclination of $29\pm 3$ degrees by modelling the jet from the radio data and \cite{carotenuto2022black} gave a upper limit of the inclination $46^{\circ}$ by the radio image. The spectral analysis of the reflection component in the X-ray spectrum determined the inclination of $\sim 29^\circ - 40^\circ$  \citep{mall2022broadband,kumar2022estimation,jia2022detailed,wu2022accretion}, which is well consistent with the jet results. The analysis of the reflection component during the second outburst in 2019 also found a inclination of $30^{\circ} - 46^{\circ}$ \citep{bhowmick2022spectral}. \cite{liu2022transitions} suggested the inclination of $\sim 30^\circ$ based on studying quasi-periodic oscillations (QPO). \cite{titarchuk2022maxi} used the scaling technique through the correlation between photon index and normalization proportional to mass accretion rate and gave the inclination of $65\pm 7$ degrees, which deviates far from other independent results, maybe due to the large uncertainties of the correlation.
%The inclination of the disk is investigated by several research: $32.9^{+4.1}_{-0.6}$ degrees \citep{mall2022broadband}, $29\pm 3$ degrees \citep{carotenuto2022modelling}, $40.5^{\circ} - 46.4^{\circ}$ \citep{bhowmick2022spectral}, $36.5\pm 1.0$ degrees \citep{kumar2022estimation}, $29.2^{+0.3}_{-0.5}$ degrees \citep{jia2022detailed}, less than $46^{\circ}$ \citep{carotenuto2022black}, $65\pm 7$ degrees \citep{titarchuk2022maxi}. 

Furthermore, both type-B and type-C QPO were found in this BHXRB \citep{alabarta2022variability,belloni2020time,liu2022transitions}. The timing analysis illustrated that there is a time lag in the 0.5-80 keV energy band \citep{jithesh2021broad} which revealed the time lag between the disk component and the corona radiation has been discovered \citep{weng2021time}. In the infrared and optical bands, the absorption lines of H, He which point to disc wind signatures have been founded \citep{panizo2022discovery}. In hard state, the Fe absorption line with high blue-shift is also found in NuSTAR data, suggesting that there are ultra-fast outflows \citep{chakraborty2021nustar,wu2022accretion} and from the Insight-HXMT data and Swift data, \cite{zhang2022peculiar} found the evidence of the optically thin disc feature. To figure out the physical properties of the disc, the reflection component is fitted by \cite{chakraborty2021nustar} which showed that the density of the disc is over $10^{20}\rm cm^{-3}$. In quiescence, \cite{carotenuto2022black}
illustrated that this source belongs to the standard (radio-loud) track.

Based on the long-term observation from Insight-HXMT in wide-band X-rays, we constrain the spin of the black hole in MAXI J1348-630 by the CF method. This paper is organized as follows: the observation and data reduction in section \ref{data reduction}, data analysis and CF method modeling results in section \ref{data analysis}, conclusion and discussion in section \ref{conclusion}.

\section{observations and data reduction}\label{data reduction}
The Insight-HXMT launched on Jun 15 2017, which is the first hard X-ray observatory of China \citep{zhang2020overview}, observed MAXI J1348-630 for over 50 days from Jan 27 2019 to Jul 29 2019. The Insight-HXMT consists of a Low Energy (LE) telescope of which the effective area is 384 $\rm cm^{2}$ in the energy range from 1 keV to 15 keV \citep{chen2020low}, a Medium Energy (ME) telescope of which the effective area is 952 $\rm cm^{2}$ in the energy range from 5 keV to 30 keV \citep{cao2020medium}, and the High Energy X-ray telescope (HE; 20-250 keV) with the effective area of 5100 cm$^2$  \citep{liu2020high}. The pileup effects of Insight-HXMT LE telescope are extremely low ($<$1\%@18000 cts/s, see \citealt{chen2020low}). The HE Net count rate from the source was lower than the background in the soft state during our observations, thus we will only use the LE and ME data for the analysis in this work. 

\begin{figure}
    \centering
    \includegraphics[scale=0.5]{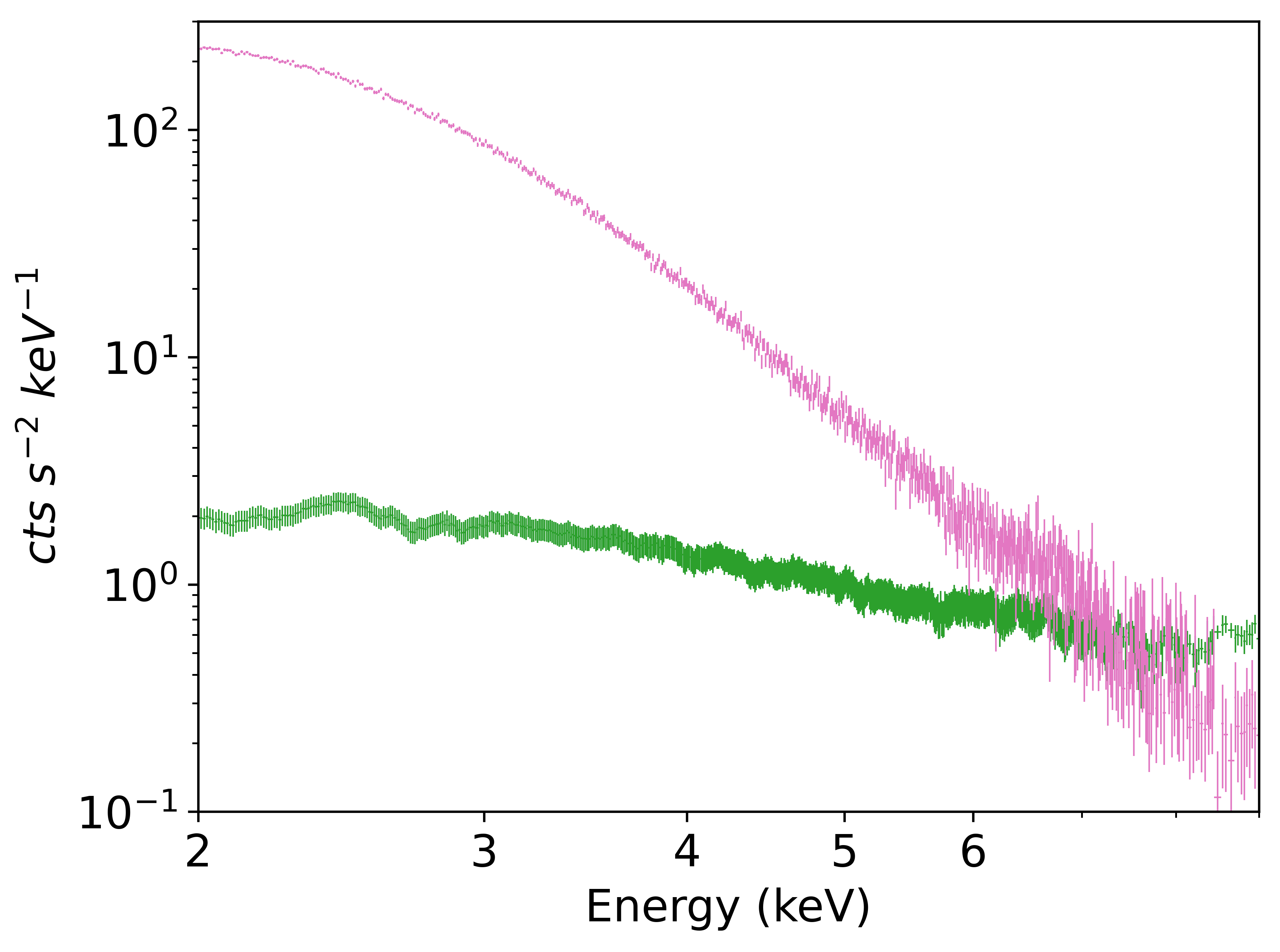}
    \caption{This figure shows the background count rate (green dots) and the source spectrum count rate (purple dots) from ObsID 301 for the LE telescope.}
    \label{background}
\end{figure}

We extract the spectra from Insight-HXMT Data using the Insight-HXMT Data Analysis software (HXMTDAS) v2.04 \footnote{http://hxmtweb.ihep.ac.cn/software.jhtml}. We jointly analyze two telescopes (LE, ME) by combining the 2 keV to 9 keV from LE due to high background noises leading to low Net count rate above 9 keV and the instrumental feature below 2 keV \citep{li2023orbit}, and the 10 keV to 20 keV from ME due to the calibration uncertainties from 21 -- 24 keV and the background noises leading to low Net count rate above $\sim$ 20 keV for the soft state \citep{cao2020medium}, and inserting two cross-normalization constants (the constant model in XSPEC), besides, we freeze the LE constant and let the ME constant vary freely. Using the recommended criteria: the elevation angle $\textgreater$ 10 degrees, the geomagnetic cutoff rigidity $\textgreater$ 8 GeV, at least 300 s away from the South Atlantic Anomaly (SAA) and the pointing offset angle $\textless 0.1^\circ$. We generate the background files by LEBKGMAP and MEBKGMAP to estimate the background count rates \citep{liao2020background}. The background count rates are relatively stable and the source spectrum count rates decrease with energy (as the example for the LE telescope shown in Fig. \ref{background}), in the soft state of this BH system, background magnitude would be much lower than the source count rate below 8 keV, and at the similar levels of the source spectrum above $\sim$ 8 keV by LE telescope and in the band of 10 -- 20 keV by ME telescope. The LE detector's readout time is around 1 ms without dead time. The ME detector dead time effect in the band of 10--20 keV is about 1\% is calculated by \cite{zhang2022peculiar}, which is within our statistical uncertainty.

\section{Data Analysis}\label{data analysis}
We investigate the X-ray spectra by the XSPEC v12.12.1\footnote{https://heasarc.gsfc.nasa.gov/xanadu/xspec/}. In all following models, the galactic neutral hydrogen column density is fixed at $8.6\times10^{21}\rm cm^{-2}$ in TBabs composition \citep{tominaga2020discovery}. We insert a cross-normalization constant between LE and ME and we freeze the constant in LE as standard as well as the constant of ME can vary freely. The confidence level of every parameter is set to 90\% to produce the error range. We analyze the spectra of the observational data illustrated in Fig. \ref{q} as red crosses. 

\subsection{Hardness intensity diagram (HID)}
\begin{figure}
    \centering
    \includegraphics[scale=0.5]{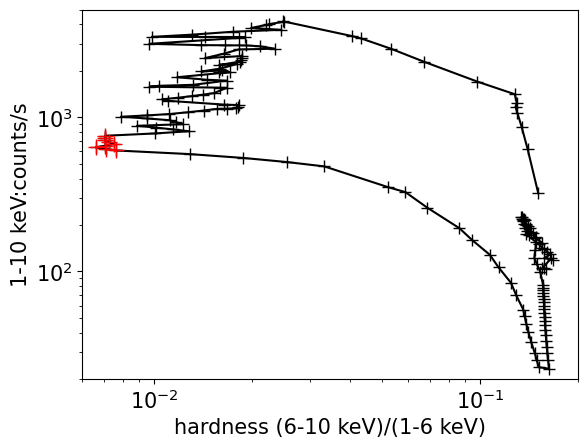}
    \caption{The HID of MAXI J1348-630 during the 2019 outburst observed by Insight-HXMT; The hardness (the count rate ratio between 6-10 keV and 1-6 keV) versus the total count rate by LE.}
    \label{q}
\end{figure}

\begin{figure}
    \centering
    \includegraphics[scale=0.5]{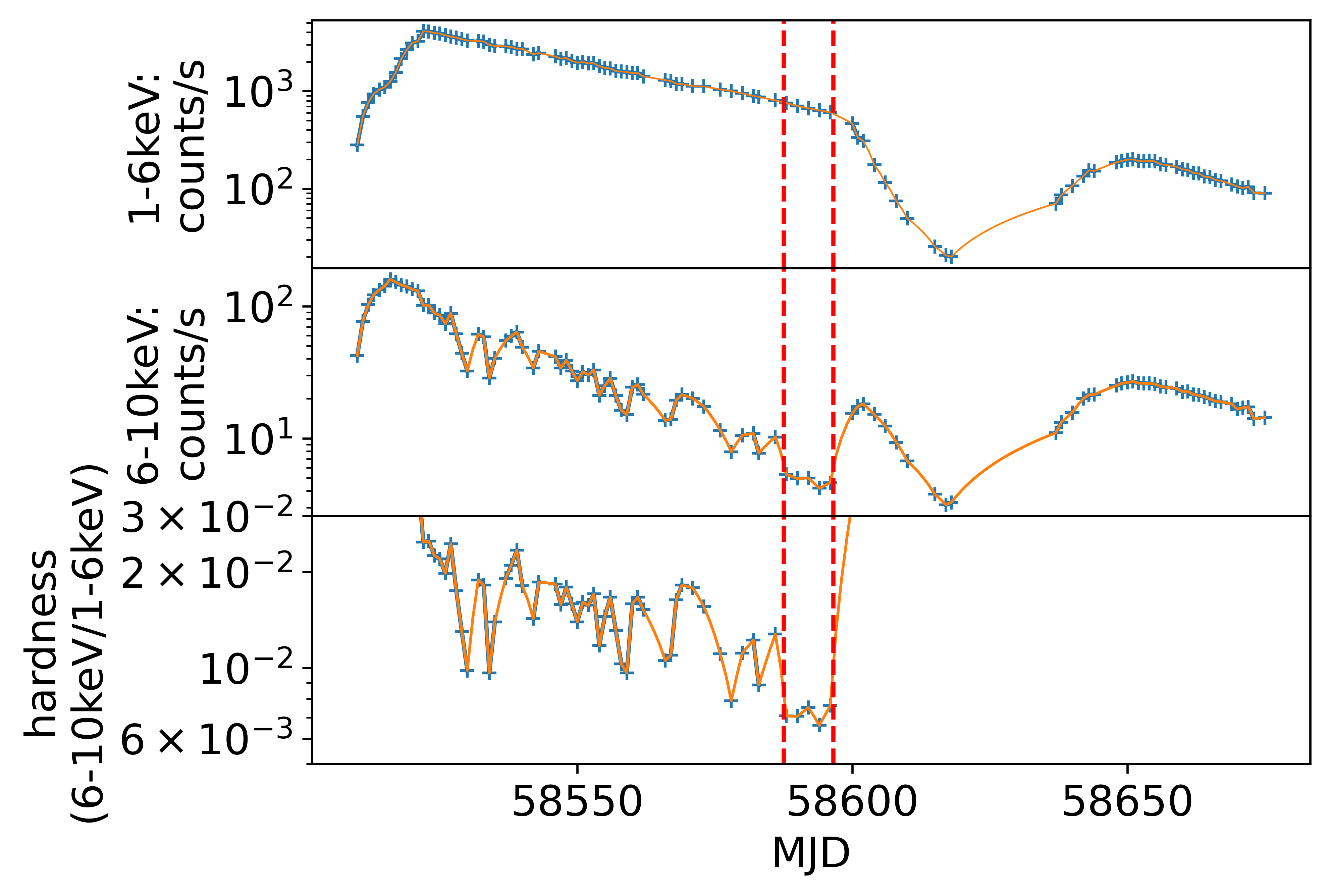}
    \caption{The light curves of MAXI J1348-630 in units of count rates observed by Insight-HXMT. The top panel illustrates the light curve of 1-6 keV photons, the middle panel for the light curve of 6-10 keV, and the bottom panel for the hardness which is the ratio between the light curve of 6-10 keV and 1-6 keV. We investigate the observational data between the red lines (MJD from 58588 to 58596).}
    \label{lc}
\end{figure}

In this outburst, the Insight-HXMT provides the pointing observations from MJD 58510 to MJD 58693, and the LE intensity versus hardness defined as the count rate between 6-10 keV and 1-6 keV are presented in hardness-intensity diagram Fig. \ref{q}, which follows a "q" pattern. 
After the peak count rate at MJD 58523, The 6-10 keV photon count rates oscillate until MJD 58588, which is illustrated in Fig. \ref{lc}. And the hardness ratio evolved from 0.02 to $\sim 8\times 10^{-3}$, between MJD 58588 and 58596, the hardness reached the lowest value around $\sim 6\times 10^{-3}$. After the softest state, the 1-6 keV photon count rate decreased by two orders of magnitude steeply and the 6-10 keV photon count rate increased by several times, and the BH system transited to the hard state again and finished the main outburst. In the following spectral analysis, we will investigate the observational data from MJD 58588 to MJD 58596. There are 11 times of observations in this period, which are illustrated in Table \ref{ob}.

\begin{table*}
\centering
\caption{Insight-HXMT observation details for MAXI J1348-630. }
\label{ob}
\begin{tabular}{lccccr} % four columns, alignment for each
    \hline
    Obs. ID & Obs. date & Obs. date & LE Exposure(s) & ME Exposure(s) & Abbreviation   \\
	    & (yyyy-mm-dd) & (MJD) &  &  \\
    \hline
    P02140206201   & 2019-04-15 & 58588 & 2933 & 2542 & 201\\
    P02140206202   & 2019-04-15 & 58588 & 1556 & 2305 & 202\\
    P02140206301   & 2019-04-17 & 58590 & 2933 & 2808 & 301\\
    P02140206302   & 2019-04-17 & 58590 & 1855 & 2333 & 302\\
    P02140206401   & 2019-04-19 & 58592 & 1731 & 1948 & 401\\
    P02140206402   & 2019-04-20 & 58593 & 947 & 1220 & 402\\
    P02140206403   & 2019-04-20 & 58593 & 1017 & 1366 & 403\\
    P02140206501   & 2019-04-21 & 58594 & 1855 & 1948 & 501\\
    P02140206601   & 2019-04-23 & 58596 & 2633 & 2762 & 601\\
    P02140206602   & 2019-04-23 & 58596 & 2095 & 2115 & 602\\
    P02140206603   & 2019-04-23 & 58596 & 1616 & 2004 & 603\\
    \hline
\end{tabular}
\end{table*}

\subsection{The spectral fittings: continuum}
We will analyze the spectral properties of Insight-HXMT observations in the softest state from MJD 58588 to 58596 when the hardness is about $6\times 10^{-3}$, which is illustrated between the red dash lines in Fig. \ref{lc}. The four models we used including phenomenological and physical models to characterize the broad-band spectra (2-9 keV from LE telescope and 10-20 keV from ME telescope) will be introduced in the following four subsections. The model Tbabs is included by all models, which considers the interstellar absorption effect with \cite{wilms2000absorption} abundances and \cite{verner1996atomic} cross-section and the column density ($N_{H}$) is fixed at the well studied value \citep{tominaga2020discovery} $8.6\times10^{21}\rm cm^{-2}$ which is also used in previous work \citep{zdziarski2022insight,cangemi2022integral,carotenuto2022modelling,jia2022detailed,chakraborty2021nustar,saha2021multi}, due to the below 2 keV photons ignored leading to insufficient constraints on the column density ($N_{H}$). 

In the continuum fitting method the parameters like the BH mass, source distance and the inclination of the system are important and affect the results. Thus, based on the previous independent measurements, the BH mass of $M_{BH} = 11\pm 2\rm M_{\odot}$, the distance of $D = 3.39\pm 0.34 \rm kpc$ \citep{lamer2021giant}, and the inclination of $i = 29.2^{+0.3}_{-0.5}\ \rm degrees$ \citep{jia2022detailed} as fiducial values are adopted in the following spectral fittings. The inclination $i = 29.2^{+0.3}_{-0.5}\ \rm degrees$ used here is also consistent with the values by several independent methods including reflection model, jet model from radio data, and QPO research \citep{mall2022broadband,kumar2022estimation,jia2022detailed,carotenuto2022modelling,liu2022transitions}. At the fiducial values, the flux (2-20 keV) is $\sim (1-2)\times 10^{-8}\rm\ erg\ s^{-1}\ cm^{-2}$, and X-ray luminosity of this source is $\sim (2-4)\times 10^{37}\rm\ erg\ s^{-1}$ during the softest state at a distance of 3.4 kpc. If taking the BH mass at 11$M_{\odot}$, the Eddington luminosity will be $1.4\times 10^{39}\rm\ erg\ s^{-1}$, then this source during the period has an accretion luminosity less than 3\% of Eddington luminosity. Furthermore, we also noticed that the mass, distance, and inclination of the black hole in MAXI J1348-630 are pretty uncertain, so in section \ref{EA}, we discuss the influence of these three key parameters (the mass, distance, and inclination of the black hole in MAXI J1348-630) on the spin determination.

%Another component constant also in all models is the cross-normalization to reconcile the calibration discrepancies between LE telescope and ME telescope (In all models and in all spectra, the LE telescope is standard).
%\begin{table*}
%\centering
%\caption{The models characterize the spectra from the Insight-HXMT and the applied energy bands}
%\label{mod}
%\begin{tabular}{lr} % four columns, alignment for each
%    \hline
%    Model & Energy bands \\
%    \hline
%    Model 1: constant $\ast$ Tbabs $\ast$ ( diskbb + powerlaw )   & 2-9 keV (LE), 10-20 keV (ME) \\
%    Model 2: constant $\ast$ Tbabs $\ast$ ( kerrbb + powerlaw )   & 2-9 keV (LE), 10-20 keV (ME) \\
%    Model 3: constant $\ast$ Tbabs $\ast$ ( kerrbb + nthcomp )   & 2-9 keV (LE), 10-20 keV (ME) \\
%    Model 4: constant $\ast$ Tbabs $\ast$ ( diskbb + nthcomp )   & 2-9 keV (LE), 10-20 keV (ME) \\
%    \hline
%\end{tabular}
%\end{table*}

\subsubsection{Model 1: constant $\ast$ Tbabs $\ast$ ( diskbb + powerlaw )}
\begin{figure}
    \centering
    \includegraphics[scale=0.5]{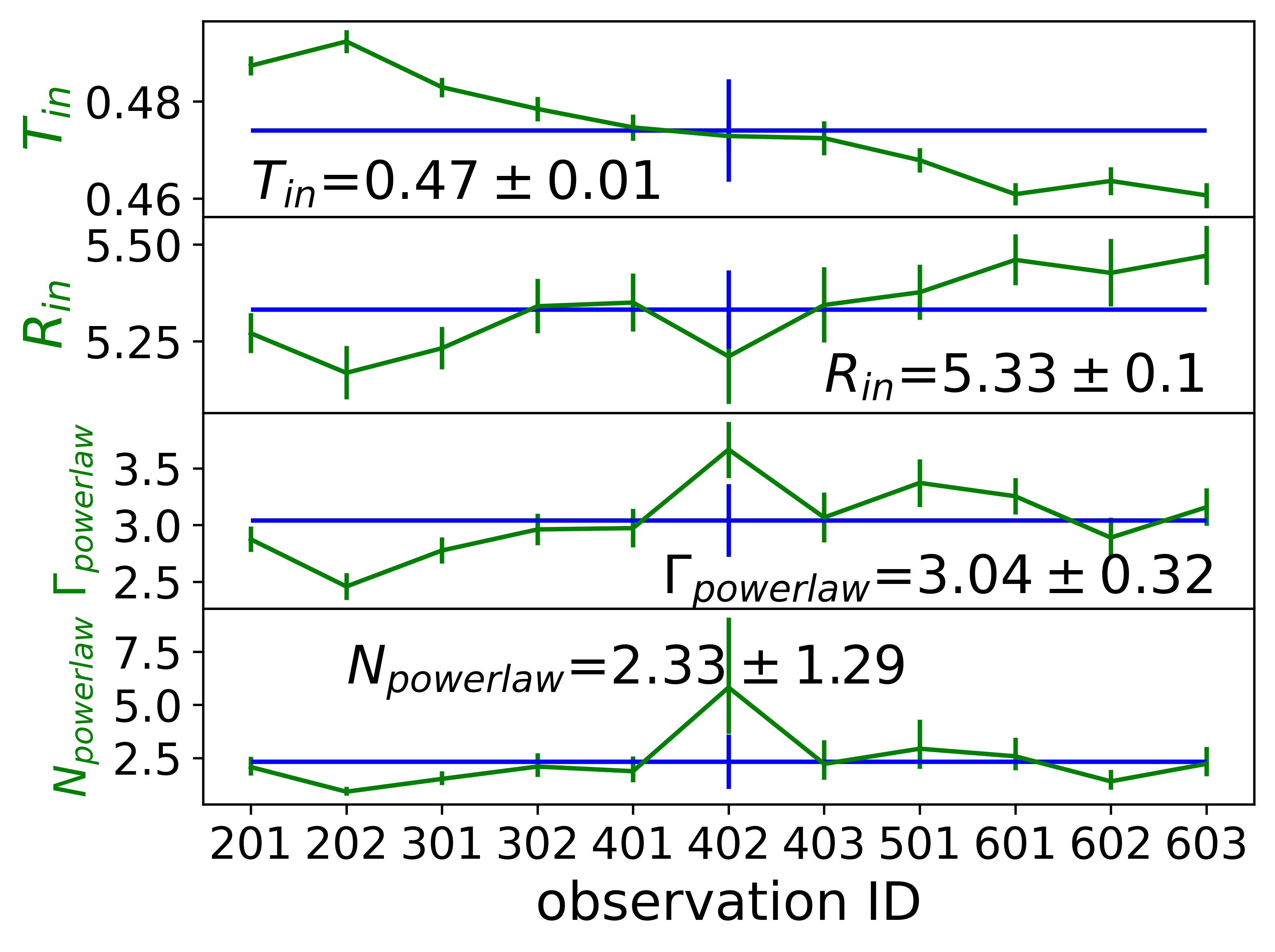}
    \caption{The fitted parameters versus the observation ID with model 1. $T_{in}$ is the inner disc temperature, $R_{in}$ for the apparent inner disc radius in units of $R_{g}$, $\Gamma_{powerlaw}$ for the photon index in powerlaw model, $N_{powerlaw}$ for the normalization of powerlaw model. The average results are shown in black font.}
    \label{M1p}
\end{figure}

\begin{figure}
    \centering
    \includegraphics[scale=0.5]{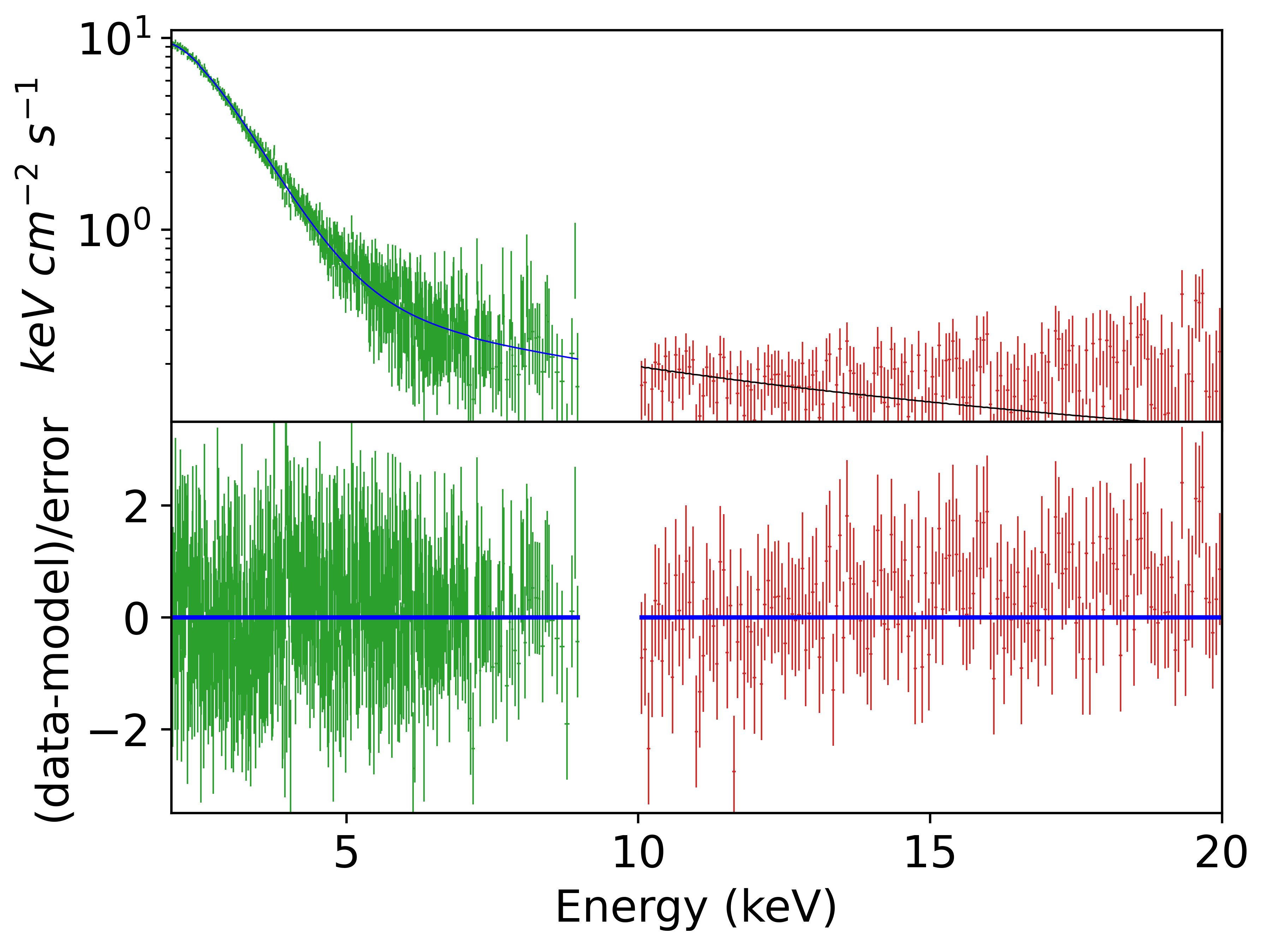}
    \caption{The spectrum and the residual for ObsID 403 with model 1 as an example. The green and red data points correspond to LE telescope and ME telescope, respectively.}
    \label{403diskpl}
\end{figure}

Firstly, the spectrum is fitted by the multicolor disc blackbody model diskbb \citep{mitsuda1984energy,makishima1986simultaneous} and a power-law component: constant $\ast$ Tbabs $\ast$ ( diskbb + powerlaw ). The results of all fitted parameters in model 1 are shown in Table \ref{M1} and Fig. \ref{M1p}, and the example of spectra and residuals are shown in Fig. \ref{403diskpl}. 
The inner disc temperature is determined at around 0.47 keV, and $\Gamma_{powerlaw}\sim 3.04$. The parameter $N_{diskbb}$ of diskbb is related to the apparent inner disc radius by the function: $N_{diskbb}=(\frac{R_{in}}{D_{10}})^{2}\cos i$, where the $R_{in}$ is the apparent inner disc radius, $D_{10}$ for the distance in units of 10 kpc, $i$ for the inclination of the system \citep{kubota1998evidence}, thus, the apparent inner disc radius $R_{in}$ is around 5.3 $R_{g}$. 

\begin{table*}
		\centering
		\caption{Spectral fitting results for Model 1: constant $\ast$ Tbabs $\ast$ ( diskbb + powerlaw )}
		\label{M1}
		\begin{tabular}{lccccccr} % four columns, alignment for each
			\hline
			Obs.($\rightarrow$) &  & 201 & 202 & 301 & 302 & 401 & 402  \\
			component($\downarrow$) & parameter($\downarrow$) &  &  &  &  \\
			\hline
			TBabs & $N_{\rm H}$ $(\times 10^{22}\rm cm^{-2})$ & $0.86^{\dag}$ & $0.86^{\dag}$ & $0.86^{\dag}$ & $0.86^{\dag}$ & $0.86^{\dag}$ & $0.86^{\dag}$ \\
			
			powerlaw & $\Gamma_{powerlaw}$ & $2.87\pm {0.11}$ & $2.46\pm {0.12}$ & $2.77\pm {0.12}$ & $2.96\pm {0.14}$ & $2.97\pm {0.17}$ & $3.67\pm {0.25}$ \\
			       & $N_{powerlaw}$ & $2.07^{+0.49}_{-0.39}$ & $0.91^{+0.23}_{-0.18}$ & $1.51^{+0.37}_{-0.29}$ & $2.09^{+0.63}_{-0.48}$ & $1.87^{+0.70}_{-0.51}$ & $5.82^{+3.29}_{-2.18}$\\
			\hline
			diskbb & $T_{in}\rm (keV)$ & $0.49\pm {0.01}$ & $0.49\pm {0.01}$ & $0.48\pm {0.01}$ & $0.48\pm {0.01}$ & $0.47\pm {0.01}$ & $0.47\pm {0.01}$\\
			 & $N_{diskbb}$  & $55486.09^{+1107.29}_{-1071.60}$ & $53350.49^{+1450.93}_{-1393.49}$ & $54676.14^{+1169.21}_{-1128.09}$ & $56966.02^{+1522.29}_{-1470.96}$ & $57170.00^{+1621.38}_{-1571.52}$ & $54230.64^{+2266.31}_{-2522.51}$\\
			 & $C_{ME}$ & $0.98\pm 0.07$ & $0.87\pm 0.06$ & $0.98\pm {0.07}$ & $1.07\pm {0.09}$ & $1.03\pm 0.11$ & $1.67^{+0.26}_{-0.22}$\\
			 & $\chi^{2}/\nu$ & $1130.63/991$ & $1106.90/991$ & $1089.43 /991$ & $1265.14/991$& $1028.30 /991$ & $1070.39 /991$\\
			 & $\chi^{2}_{\nu}$ & $1.14 $ & $1.12 $ & $1.10 $ & $1.28 $ & $1.04 $ & $1.08 $\\
			\hline
			Obs.($\rightarrow$) &  & 403 & 501 & 601 & 602 & 603  \\
			component($\downarrow$) & parameter($\downarrow$) &  &  &  &  \\
			\hline
			TBabs & $N_{\rm H}$ $(\times 10^{22}\rm cm^{-2})$ & $0.86^{\dag}$ & $0.86^{\dag}$ & $0.86^{\dag}$ & $0.86^{\dag}$ & $0.86^{\dag}$  \\
			
			powerlaw & $\Gamma_{powerlaw}$ & $3.07\pm {0.22}$ & $3.37\pm {0.21}$ & $3.25\pm {0.16}$ & $2.89\pm {0.18}$ & $3.16\pm {0.17}$  \\
			       & $N_{powerlaw}$ & $2.22^{+1.12}_{-0.74}$ & $2.93^{+1.37}_{-0.95}$ & $2.58^{+0.88}_{-0.66}$ & $1.40^{+0.55}_{-0.39}$ & $2.22^{+0.80}_{-0.58}$\\
			\hline
			diskbb & $T_{in}\rm (keV)$ & $0.47\pm {0.01}$ & $0.47\pm 0.01$ & $0.46\pm {0.01}$ & $0.46\pm {0.01}$& $0.46\pm {0.01}$ \\
			 & $N_{diskbb}$ & $57036.20^{+2117.68}_{-2032.50}$ & $57737.78^{+1556.00}_{-1507.49}$ & $59552.11^{+1465.69}_{-1420.43}$ & $58819.82^{+1926.28}_{-1847.55}$ & $59787.99^{+1697.83}_{-1631.34}$\\
			 & $C_{ME}$ & $1.03^{+0.15}_{-0.13}$ & $1.19^{+0.15}_{-0.16}$ & $1.21^{+0.12}_{-0.13}$ & $1.05^{+0.11}_{-0.13}$& $1.30^{+0.13}_{-0.14}$ \\
			 & $\chi^{2}/\nu$ & $943.36 /991$ & $995.21 /991$ & $1018.32 /991$ & $1008.91 /991$ & $1024.60 /991$ \\
			 & $\chi^{2}_{\nu}$ & $0.95 $ & $1.00 $ & $1.03 $ & $1.02 $& $1.03 $ \\
			\hline
		\end{tabular}
	    \end{table*}

\subsubsection{Model 2: constant $\ast$ Tbabs $\ast$ ( kerrbb + powerlaw )}
\begin{figure}
    \centering
    \includegraphics[scale=0.5]{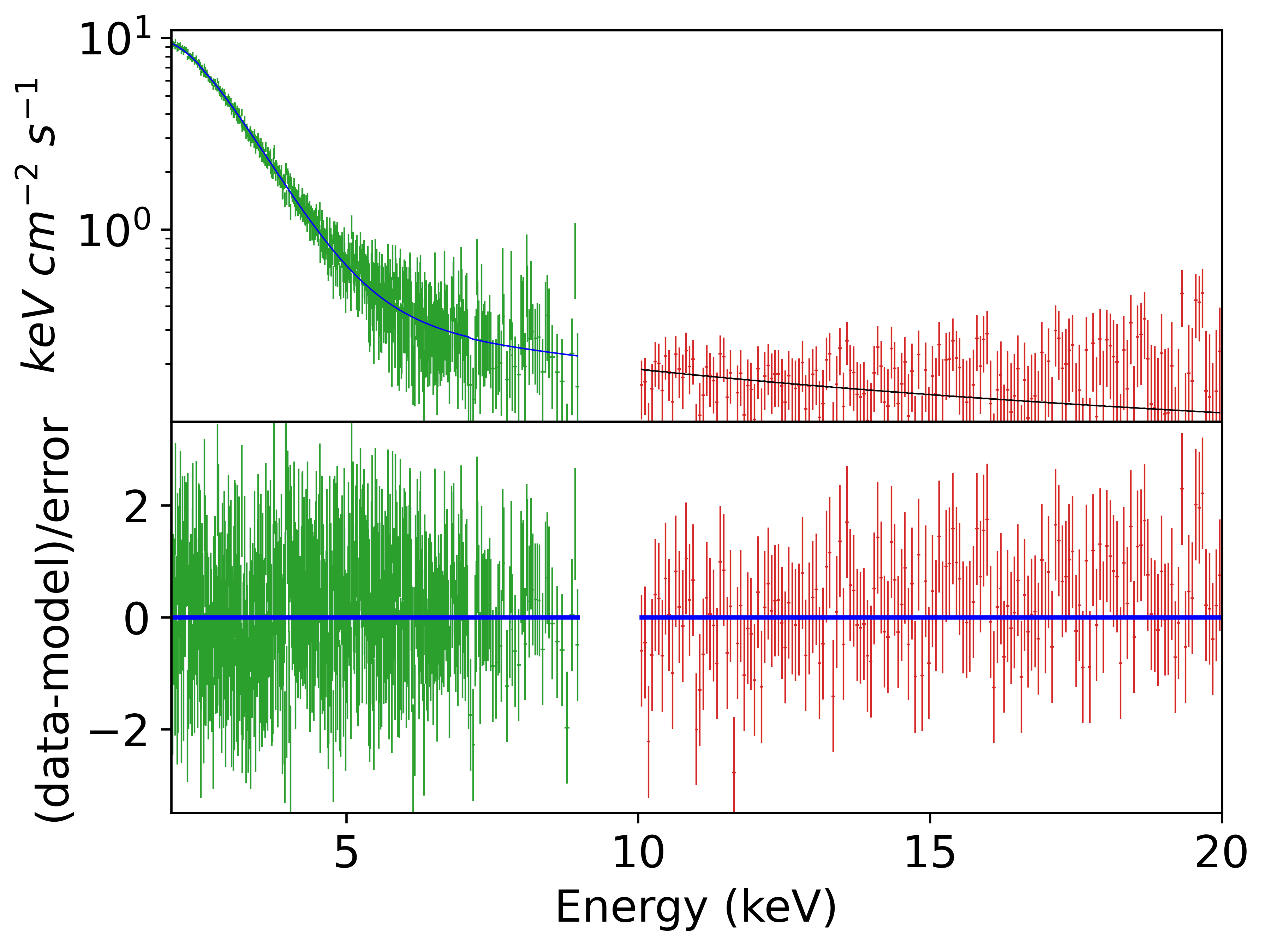}
    \caption{The spectrum and the residual for ObsID 403 with model 2 as an example. The green and red data points correspond to LE telescope and ME telescope, respectively.}
    \label{403pl}
\end{figure}

\begin{figure}
    \centering
    \includegraphics[scale=0.5]{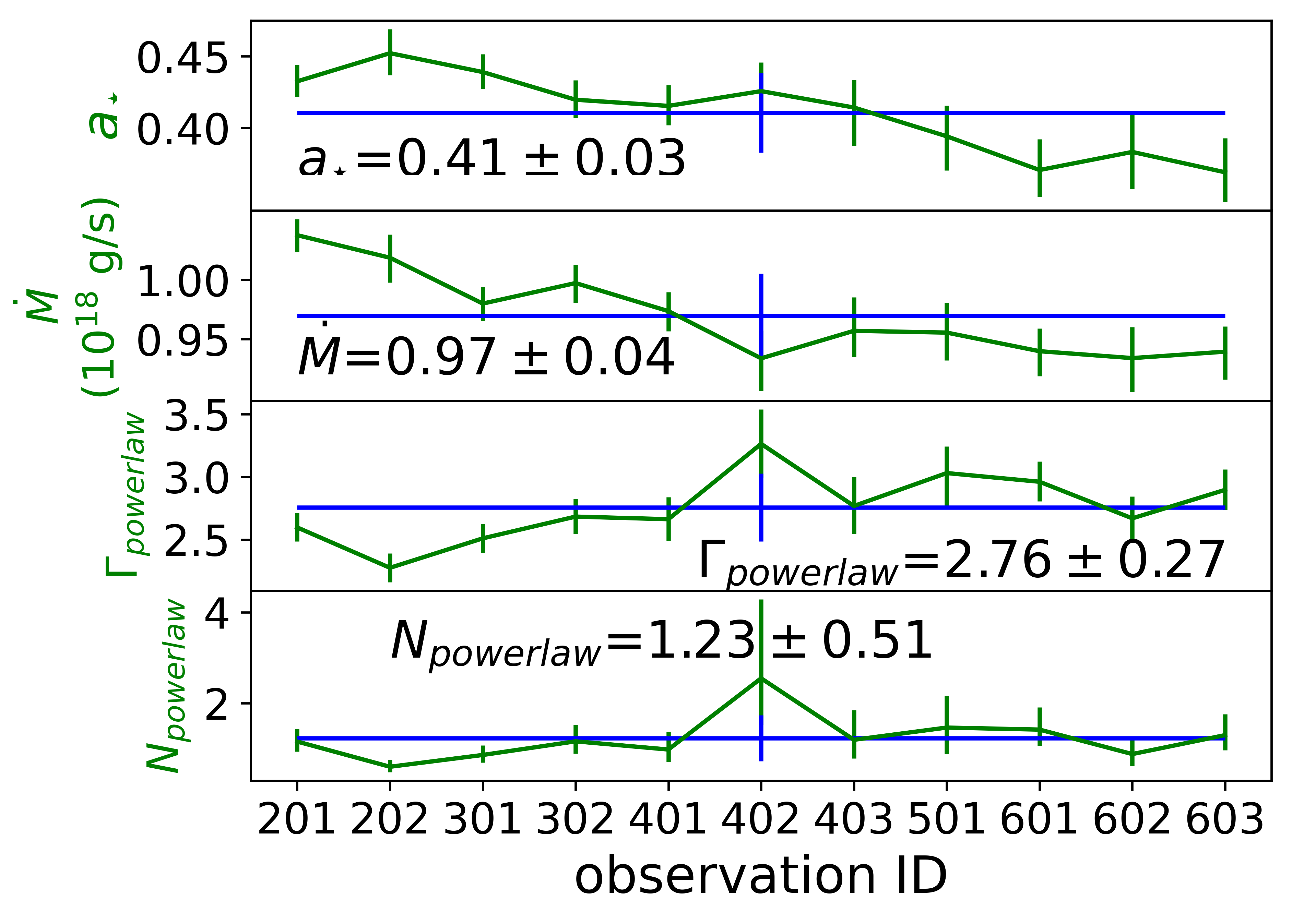}
    \caption{The fitted parameters versus the observation ID with model 2. $a_{\star}$ is the spin of black hole, $\dot{M}$ for the effective mass accretion rate in units of $10^{18} g/s$, $\Gamma_{powerlaw}$ for the photon index in powerlaw model, $N_{powerlaw}$ for the normalization of powerlaw model. The average results are shown in black font.}
    \label{M2p}
\end{figure}

Then, we take the kerrbb model\citep{li2005multitemperature} to replace the diskbb component to determine the black hole's spin $a_{\star}$. We fix the hardening factor at 1.6 \citep{shimura1995spectral,shafee2005estimating} and use the fiducial values: the BH mass of $M_{BH} = 11\pm 2\ \rm M_{\odot}$, the distance of $D = 3.39\pm 0.34 \ \rm kpc$ \citep{lamer2021giant}, and the inclination of $i = 29.2^{+0.3}_{-0.5}\rm degree$ \citep{jia2022detailed}, besides, freeze the normalization of kerrbb. The results of all fitted parameters are shown in Table \ref{M2} and Fig. \ref{M2p}. The example of spectral data and residuals are shown in Fig. \ref{403pl}.
Thus, we get the spin of the black hole in MAXI J11348-630 $a_{\star}\sim 0.41\pm 0.03$ in model 2 and the effective mass accretion rate is $0.97\times 10^{18}\rm g/s$.

\begin{table*}
		\centering
		\caption{Spectral fitting results for Model 2: constant $\ast$ Tbabs $\ast$ ( kerrbb + powerlaw )}
		\label{M2}
		\begin{tabular}{lccccccr} % four columns, alignment for each
			\hline
			Obs.($\rightarrow$) &  & 201 & 202 & 301 & 302 & 401 & 402  \\
			component($\downarrow$) & parameter($\downarrow$) &  &  &  &  \\
			\hline
			TBabs & $N_{\rm H}$ $(\times 10^{22}\rm cm^{-2})$ & $0.86^{\dag}$ & $0.86^{\dag}$ & $0.86^{\dag}$ & $0.86^{\dag}$ & $0.86^{\dag}$ & $0.86^{\dag}$ \\
			
			powerlaw & $\Gamma_{powerlaw}$ & $2.60\pm {0.11}$ & $2.27\pm {0.12}$ & $2.51\pm {0.12}$ & $2.68\pm {0.14}$ & $2.66\pm {0.17}$ & $3.26\pm {0.27}$ \\
			       & $N_{powerlaw}$ & $1.16^{+0.28}_{-0.22}$ & $0.60^{+0.15}_{-0.12}$ & $0.86^{+0.21}_{-0.17}$ & $1.16^{+0.36}_{-0.27}$ & $0.98^{+0.39}_{-0.27}$ & $2.55^{+1.74}_{-1.02}$\\
			\hline
			kerrbb & $a_{\star}$ & $0.43\pm {0.01}$ & $0.45\pm {0.02}$ & $0.44\pm {0.01}$ & $0.42\pm {0.01}$ & $0.42\pm {0.01}$ & $0.43\pm {0.02}$\\
			 & $\dot{M}$ ($10^{18} g/s$) & $1.04\pm{0.01}$ & $1.02\pm {0.02}$ & $0.98\pm {0.01}$ & $1.00\pm {0.02}$ & $0.97\pm {0.02}$ & $0.93\pm {0.02}$\\
			 & $C_{ME}$ & $0.89^{+0.06}_{-0.07}$ & $0.84^{+0.06}_{-0.07}$ & $0.90^{+0.06}_{-0.07}$ & $0.98^{+0.08}_{-0.09}$ & $0.93^{+0.09}_{-0.11}$ & $1.48^{+0.21}_{-0.24}$\\
			 & $\chi^{2}/\nu$ & $1043.83/991$ & $1031.93/991$ & $1000.12/991$ & $1195.61/991$& $981.7/991$ & $1044.89/991$\\
			 & $\chi^{2}_{\nu}$ & $1.05$ & $1.04$ & $1.01$ & $1.21$ & $0.99$ & $1.05$\\
			\hline
			Obs.($\rightarrow$) &  & 403 & 501 & 601 & 602 & 603  \\
			component($\downarrow$) & parameter($\downarrow$) &  &  &  &  \\
			\hline
			TBabs & $N_{\rm H}$ $(\times 10^{22}\rm cm^{-2})$ & $0.86^{\dag}$ & $0.86^{\dag}$ & $0.86^{\dag}$ & $0.86^{\dag}$ & $0.86^{\dag}$  \\
			
			powerlaw & $\Gamma_{powerlaw}$ & $2.77^{+0.23}_{-0.22}$ & $3.03^{+0.21}_{-0.26}$ & $2.96\pm {0.16}$ & $2.67\pm {0.17}$ & $2.90\pm {0.16}$  \\
			       & $N_{powerlaw}$ & $1.19^{+0.65}_{-0.41}$ & $1.46^{+0.58}_{-0.70}$ & $1.42^{+0.36}_{-0.49}$ & $0.88^{+0.26}_{-0.34}$ & $1.30^{+0.46}_{-0.33}$\\
			\hline
			kerrbb & $a_{\star}$ & $0.41\pm {0.03}$ & $0.39\pm {0.02}$ & $0.37\pm {0.02}$ & $0.38\pm {0.03}$& $0.37\pm {0.02}$ \\
			 & $\dot{M}$ ($10^{18} g/s$)& $0.96\pm {0.02}$ & $0.96\pm {0.02}$ & $0.94\pm {0.02}$ & $0.93\pm {0.03}$ & $0.94\pm {0.02}$\\
			 & $C_{ME}$ & $0.93^{+0.14}_{-0.12}$ & $1.06^{+0.15}_{-0.14}$ & $1.09^{+0.12}_{-0.11}$ & $0.98^{+0.12}_{-0.11}$& $1.20^{+0.13}_{-0.12}$ \\
			 & $\chi^{2}/\nu$ & $927.02/991$ & $974.47/991$ & $973.57/991$ & $974.41/991$ & $966.05/991$ \\
			 & $\chi^{2}_{\nu}$ & $0.94$ & $0.98$ & $0.98$ & $0.98$& $0.97$ \\
			\hline
		\end{tabular}
	    \end{table*}

\subsubsection{Model 3: constant $\ast$ Tbabs $\ast$ ( kerrbb + nthcomp )}
\begin{figure}
    \centering
    \includegraphics[scale=0.5]{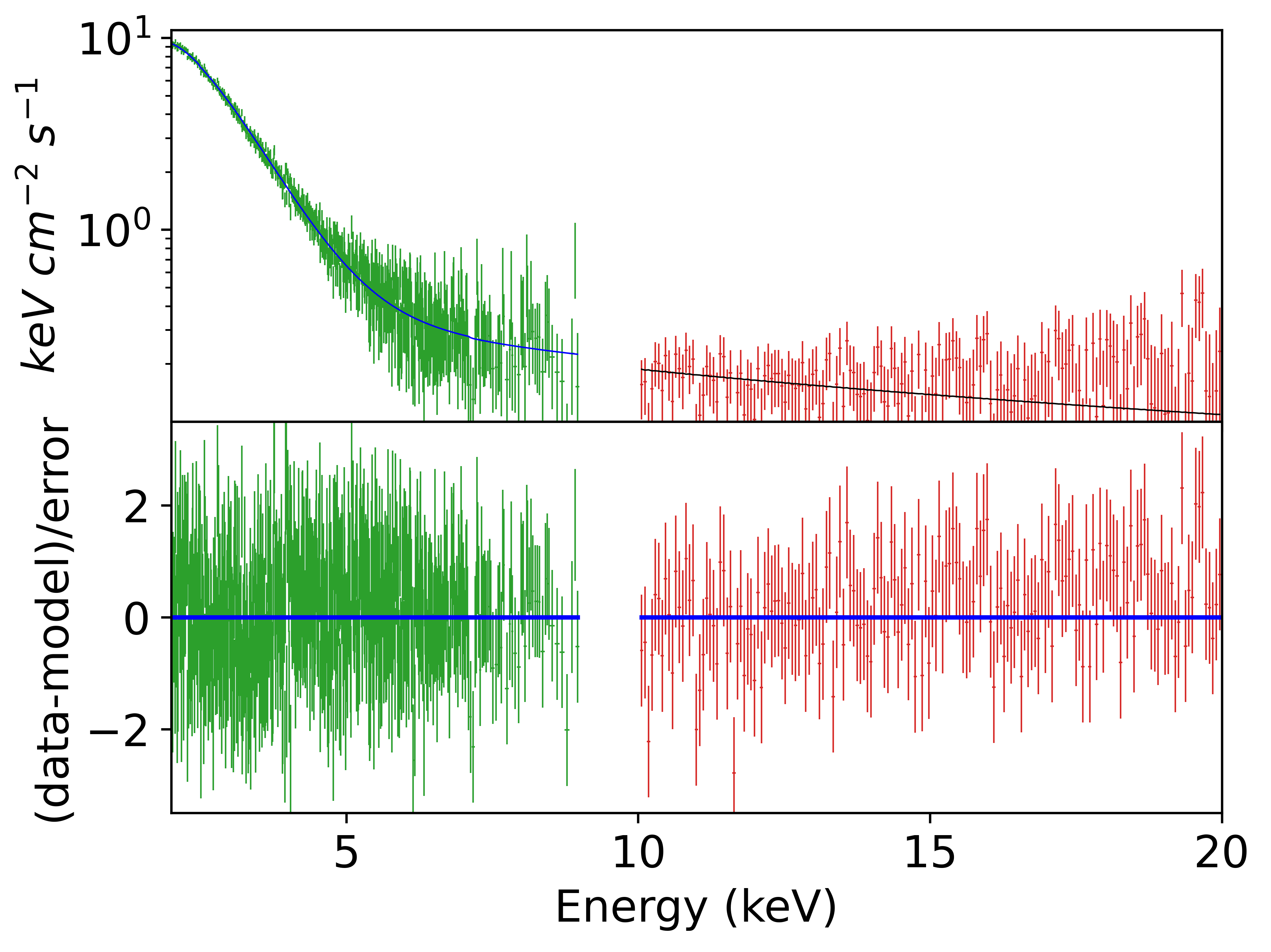}
    \caption{The spectrum and the residual for ObsID 403 with model 3 as an example. The green and red data points correspond to LE telescope and ME telescope, respectively.}
    \label{403nthpl}
\end{figure}

\begin{figure}
    \centering
    \includegraphics[scale=0.5]{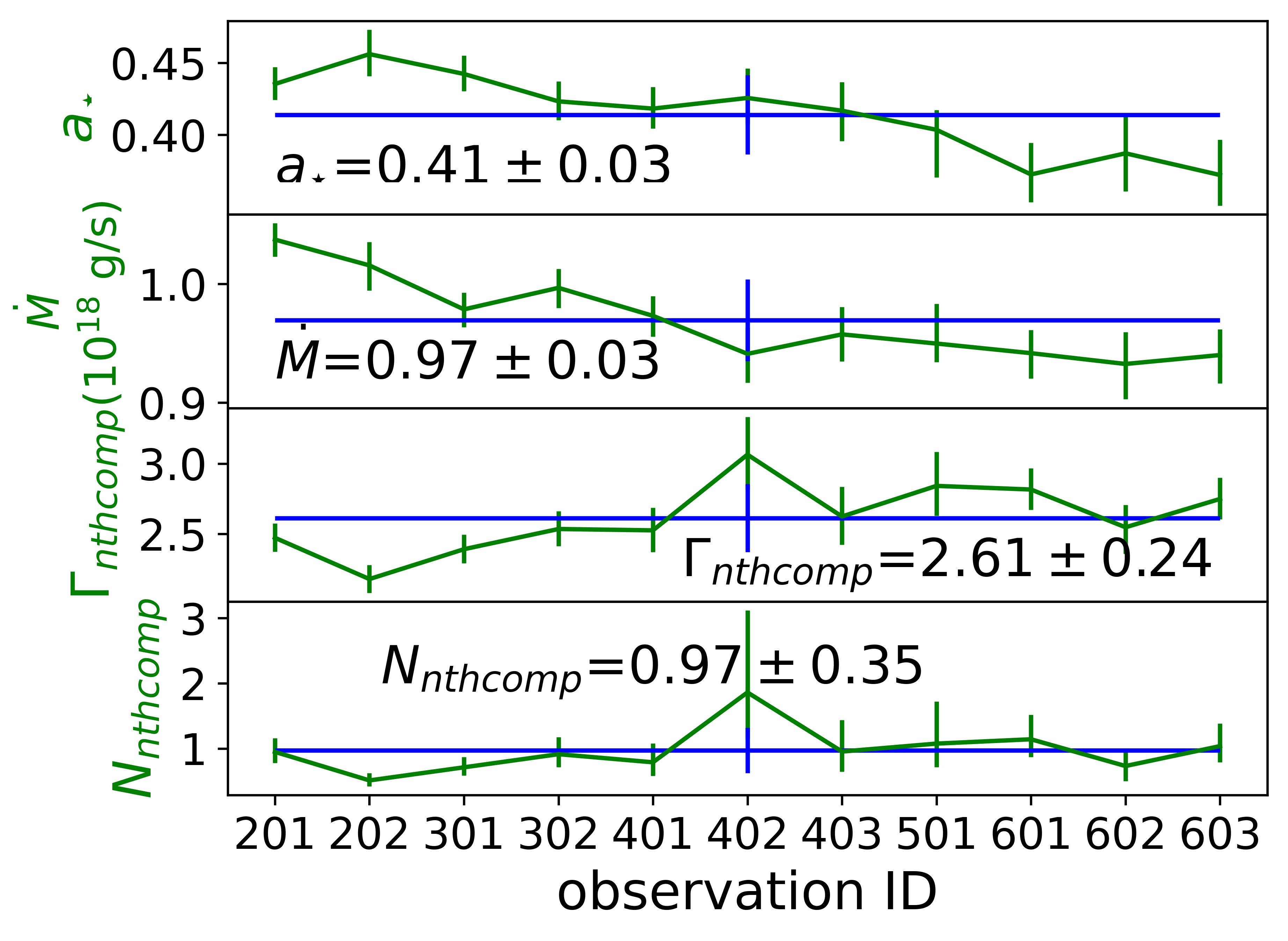}
    \caption{The fitted parameters versus the observation ID with model 3. $a_{\star}$ is the spin of the black hole, $\dot{M}$ for the effective mass accretion rate in units of $10^{18} g/s$, $\Gamma_{nthcomp}$ for the photon index in nthcomp model, $N_{nthcomp}$ for the normalization of nthcomp model. The average results are shown in black font.}
    \label{M3p}
\end{figure}

For investigating the influence of the different powerlaw components on the results of disc components, we replace the powerlaw with the nthcomp (the spectra from Comptonization by the thermal electrons \citep{zdziarski1996broad,zdziarski2020spectral,niedzwiecki2019improved}). Due to the energy range which we study can't reach the cutoff energy, we freeze the temperature of electrons at 30 keV \citep{jia2022detailed,chakraborty2021nustar} which is the electron temperature found by the relxillCp reflection model\citep{garcia2015estimating,dauser2014role}, and freeze the seed photon temperature at 0.1 keV. The results of all fitted parameters are shown in Table \ref{M3} and Fig. \ref{M3p}. The example of data and residuals are shown in Fig. \ref{403nthpl}.
Thus, we get the spin of the black hole in MAXI J11348-630 $a_{\star}\sim 0.41\pm 0.03$ in model 2 and the effective mass accretion rate $\sim 0.97\times 10^{18}\rm g/s$.

\begin{table*}
		\centering
		\caption{Spectral fitting results for Model 3: constant $\ast$ Tbabs $\ast$ ( kerrbb + nthcomp )}
		\label{M3}
		\begin{tabular}{lccccccr} % four columns, alignment for each
			\hline
			Obs.($\rightarrow$) &  & 201 & 202 & 301 & 302 & 401 & 402  \\
			component($\downarrow$) & parameter($\downarrow$) &  &  &  &  \\
			\hline
			TBabs & $N_{\rm H}$ $(\times 10^{22}\rm cm^{-2})$ & $0.86^{\dag}$ & $0.86^{\dag}$ & $0.86^{\dag}$ & $0.86^{\dag}$ & $0.86^{\dag}$ & $0.86^{\dag}$ \\
			
			nthcomp & $\Gamma_{nthcomp}$ & $2.47\pm {0.10}$ & $2.18\pm {0.10}$ & $2.39\pm {0.10}$ & $2.53^{+0.13}_{-0.12}$ & $2.53^{+0.15}_{-0.16}$ & $3.06^{+0.25}_{-0.27}$ \\
			       & $N_{nthcomp}$ & $0.95^{+0.17}_{-0.21}$ & $0.51^{+0.09}_{-0.11}$ & $0.71^{+0.16}_{-0.13}$ & $0.91^{+0.26}_{-0.20}$ & $0.79^{+0.29}_{-0.21}$ & $1.86^{+1.26}_{-0.71}$\\
			\hline
			kerrbb & $a_{\star}$ & $0.44\pm {0.01}$ & $0.46\pm {0.02}$ & $0.44\pm {0.01}$ & $0.42\pm {0.01}$ & $0.42\pm {0.01}$ & $0.43\pm {0.02}$\\
			 & $\dot{M}$ ($10^{18} g/s$) & $1.04\pm {0.01}$ & $1.02\pm {0.02}$ & $0.98\pm {0.01}$ & $1.00\pm {0.02}$ & $0.97\pm {0.02}$ & $0.94\pm {0.02}$\\
			 & $C_{ME}$ & $0.87\pm {0.06}$ & $0.82\pm {0.06}$ & $0.88^{+0.07}_{-0.06}$ & $0.96^{+0.09}_{-0.08}$ & $0.91^{+0.10}_{-0.09}$ & $1.44^{+0.24}_{-0.20}$\\
			 & $\chi^{2}/\nu$ & $1065.73 /991$ & $1043.14 /991$ & $1016.76 /991$ & $1216.24 /991$& $996.94 /991$ & $1055.53 /991$\\
			 & $\chi^{2}_{\nu}$ & $1.08 $ & $1.05 $ & $1.03 $ & $1.23 $ & $1.01 $ & $1.07 $\\
			\hline
			Obs.($\rightarrow$) &  & 403 & 501 & 601 & 602 & 603  \\
			component($\downarrow$) & parameter($\downarrow$) &  &  &  &  \\
			\hline
			TBabs & $N_{\rm H}$ $(\times 10^{22}\rm cm^{-2})$ & $0.86^{\dag}$ & $0.86^{\dag}$ & $0.86^{\dag}$ & $0.86^{\dag}$ & $0.86^{\dag}$  \\
			
			nthcomp & $\Gamma_{nthcomp}$ & $2.62^{+0.20}_{-0.21}$ & $2.84^{+0.21}_{-0.24}$ & $2.82^{+0.14}_{-0.15}$ & $2.55^{+0.19}_{-0.16}$ & $2.75^{+0.14}_{-0.15}$  \\
			       & $N_{nthcomp}$ & $0.95^{+0.48}_{-0.31}$ & $1.07^{+0.64}_{-0.36}$ & $1.14^{+0.37}_{-0.27}$ & $0.73^{+0.26}_{-0.23}$ & $1.04^{+0.34}_{-0.25}$\\
			\hline
			kerrbb & $a_{\star}$ & $0.42\pm {0.02}$ & $0.40^{+0.01}_{-0.03}$ & $0.37\pm {0.02}$ & $0.39\pm {0.03}$& $0.37\pm {0.02}$ \\
			 & $\dot{M}$ ($10^{18} g/s$)& $0.96\pm {0.02}$ & $0.95\pm {0.03}$ & $0.94\pm {0.02}$ & $0.93\pm {0.03}$ & $0.94\pm {0.02}$\\
			 & $C_{ME}$ & $0.90^{+0.14}_{-0.12}$ & $1.03^{+0.13}_{-0.16}$ & $1.07^{+0.12}_{-0.10}$ & $0.97^{+0.12}_{-0.10}$& $1.17^{+0.13}_{-0.11}$ \\
			 & $\chi^{2}/\nu$ & $926.16 /991$ & $980.22 /991$ & $989.21 /991$ & $993.09 /991$ & $997.22 /991$ \\
			 & $\chi^{2}_{\nu}$ & $0.93 $ & $0.99 $ & $1.00 $ & $1.00$& $1.01 $ \\
			\hline
		\end{tabular}
	    \end{table*}

\subsubsection{Model 4: constant $\ast$ Tbabs $\ast$ ( diskbb + nthcomp )}
\begin{figure}
    \centering
    \includegraphics[scale=0.5]{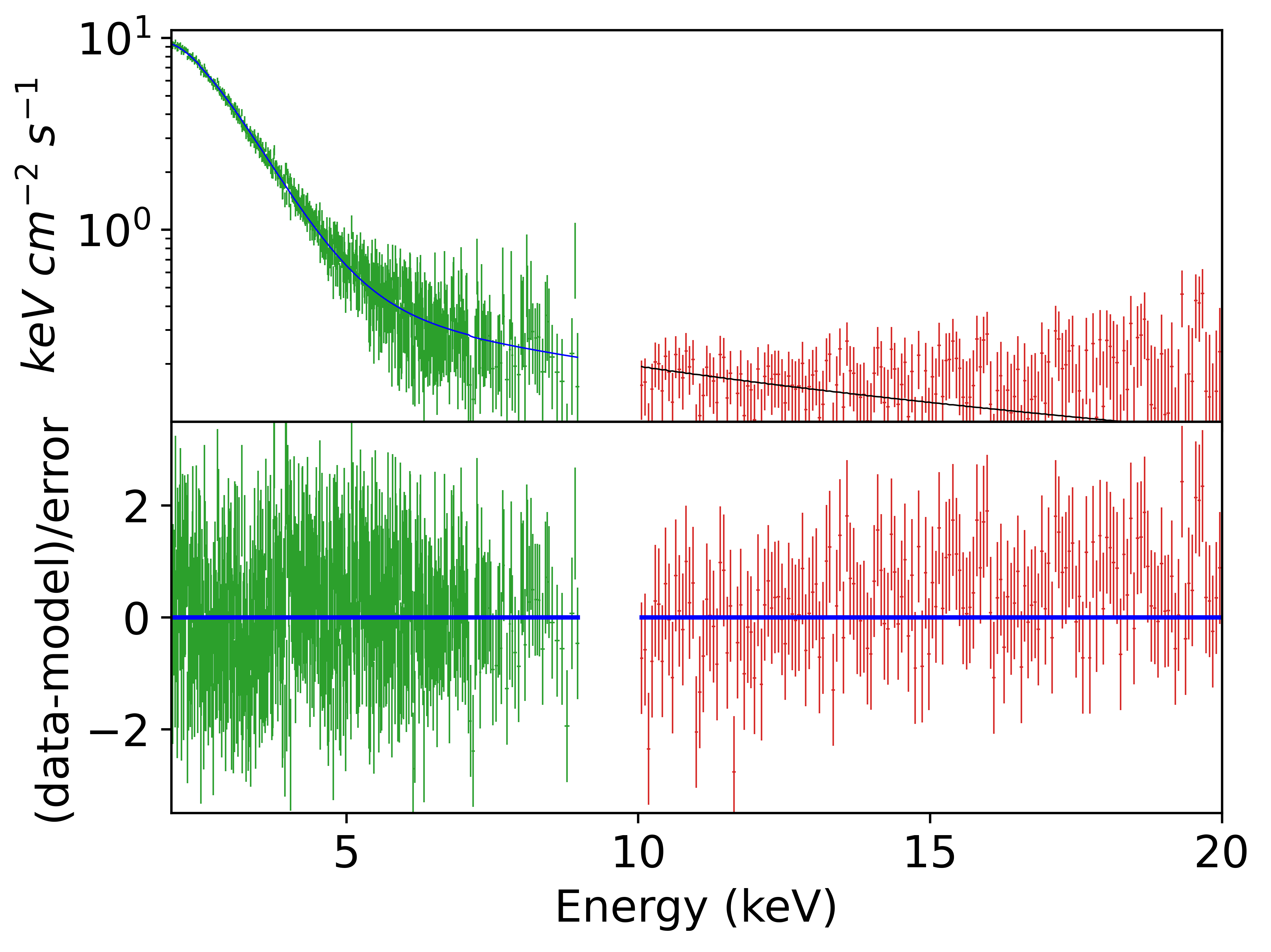}
    \caption{The spectrum and the residual for ObsID 403 with model 4 as an example. The green and red data points correspond to LE telescope and ME telescope, respectively.}
    \label{403disknthpl}
\end{figure}

\begin{figure}
    \centering
    \includegraphics[scale=0.5]{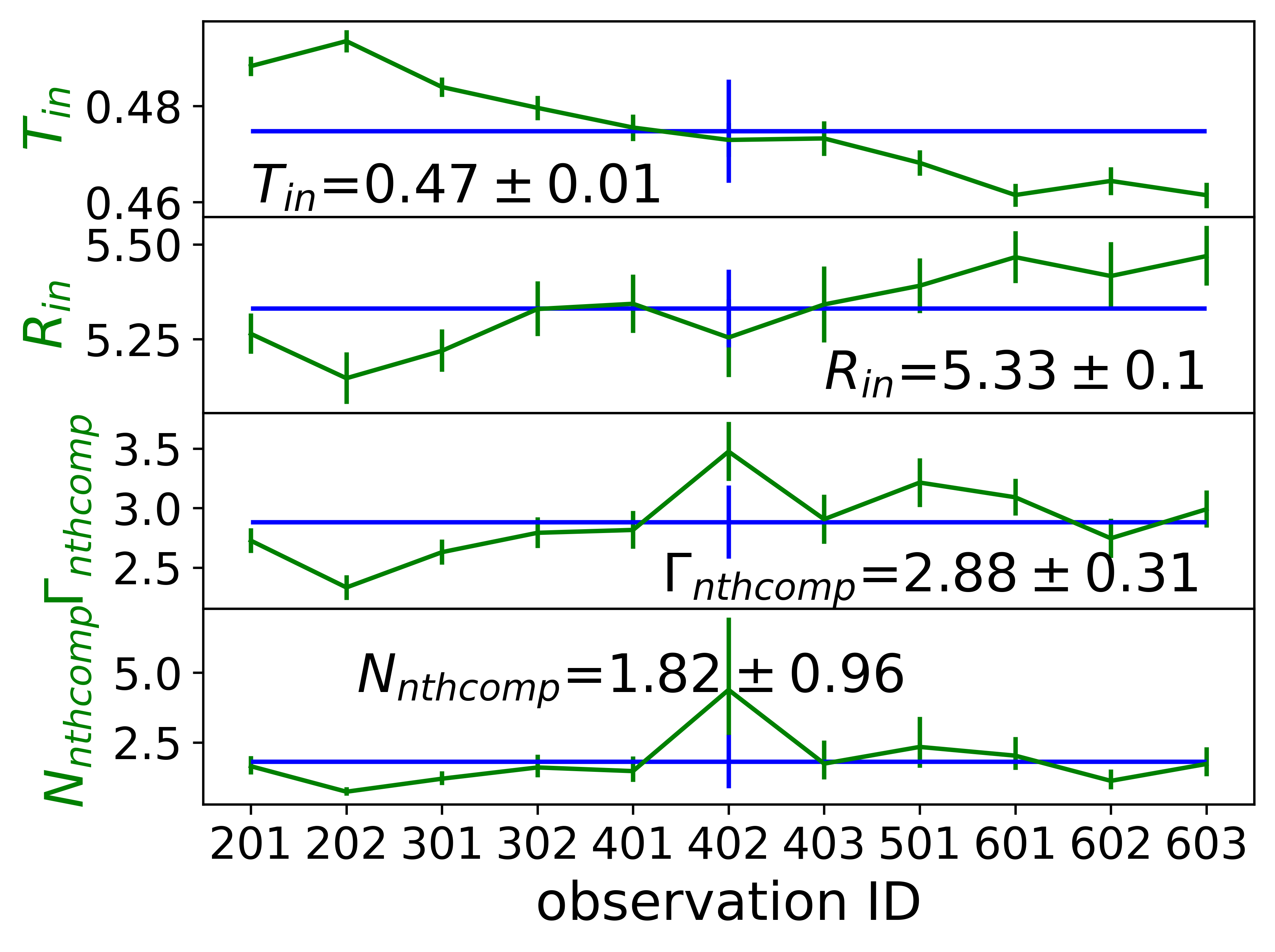}
    \caption{The fitted parameters versus the observation ID with model 4. $T_{in}$ is the inner disc temperature, $R_{in}$ for the apparent inner disc radius in units of $R_{g}$, $\Gamma_{nthcomp}$ for the photon index in nthcomp model, $N_{nthcomp}$ for the normalization of nthcomp model. The average results are shown in black font.}
    \label{M4p}
\end{figure}

Finally, we test the multicolor disc blackbody model diskbb with nthcomp component. The nthcomp component is set the same as model 3. The results of all fitted parameters are shown in Table \ref{M4} and Fig. \ref{M4p}. The example of data and residuals are shown in Fig. \ref{403disknthpl}.
The results of diskbb in model 4 are quite close to model 1 results: the apparent inner disc radius $R_{in}\sim 5.3 R_{g}$ and the inner disc temperature at around 0.47 keV.

\begin{table*}
		\centering
		\caption{Spectral fitting results for Model 4: constant $\ast$ Tbabs $\ast$ ( diskbb + nthcomp )}
		\label{M4}
		\begin{tabular}{lccccccr} % four columns, alignment for each
			\hline
			Obs.($\rightarrow$) &  & 201 & 202 & 301 & 302 & 401 & 402  \\
			component($\downarrow$) & parameter($\downarrow$) &  &  &  &  \\
			\hline
			TBabs & $N_{\rm H}$ $(\times 10^{22}\rm cm^{-2})$ & $0.86^{\dag}$ & $0.86^{\dag}$ & $0.86^{\dag}$ & $0.86^{\dag}$ & $0.86^{\dag}$ & $0.86^{\dag}$ \\
			
			nthcomp & $\Gamma_{nthcomp}$ & $2.73\pm {0.10}$ & $2.33\pm {0.10}$ & $2.63\pm {0.11}$ & $2.79\pm {0.13}$ & $2.82\pm {0.16}$ & $3.47\pm {0.25}$ \\
			       & $N_{nthcomp}$ & $1.65^{+0.37}_{-0.30}$ & $0.74^{+0.17}_{-0.14}$ & $1.20^{+0.28}_{-0.22}$ & $1.60^{+0.46}_{-0.35}$ & $1.47^{+0.53}_{-0.38}$ & $4.38^{+2.59}_{-1.63}$\\
			\hline
			diskbb & $T_{in}\rm (keV)$ & $0.49\pm {0.01}$ & $0.49\pm {0.01}$ & $0.48\pm {0.01}$ & $0.48\pm {0.01}$ & $0.48\pm {0.01}$ & $0.47\pm {0.01}$\\
			 & $N_{diskbb}$  & $55348.54^{+1132.25}_{-1098.08}$ & $52913.45^{+1429.77}_{-1379.49}$ & $54410.04^{+1184.22}_{-1144.24}$ & $56729.49^{+1561.19}_{-1498.12}$ & $57023.17^{+1668.25}_{-1610.38}$ & $55142.53^{+2137.58}_{-2157.72}$\\
			 & $C_{ME}$ & $0.96^{+0.07}_{-0.06}$ & $0.85^{+0.07}_{-0.06}$ & $0.95\pm {0.07}$ & $1.04^{+0.09}_{-0.08}$ & $1.01^{+0.11}_{-0.10}$ & $1.64^{+0.26}_{-0.22}$\\
			 & $\chi^{2}/\nu$ & $1153.01 /991$ & $1120.52 /991$ & $1109.78 /991$ & $1285.89 /991$& $1040.74 /991$ & $1081.20 /991$\\
			 & $\chi^{2}_{\nu}$ & $1.16 $ & $1.13 $ & $1.12 $ & $1.30 $ & $1.05 $ & $1.09 $\\
			\hline
			Obs.($\rightarrow$) &  & 403 & 501 & 601 & 602 & 603  \\
			component($\downarrow$) & parameter($\downarrow$) &  &  &  &  \\
			\hline
			TBabs & $N_{\rm H}$ $(\times 10^{22}\rm cm^{-2})$ & $0.86^{\dag}$ & $0.86^{\dag}$ & $0.86^{\dag}$ & $0.86^{\dag}$ & $0.86^{\dag}$  \\
			
			nthcomp & $\Gamma_{nthcomp}$ & $2.91\pm {0.20}$ & $3.21\pm {0.20}$ & $3.09\pm {0.16}$ & $2.75\pm 0.16$ & $2.99\pm {0.16}$  \\
			       & $N_{nthcomp}$ & $1.74^{+0.84}_{-0.56}$ & $2.34^{+1.08}_{-0.74}$ & $2.03^{+0.68}_{-0.50}$ & $1.12^{+0.42}_{-0.30}$ & $1.73^{+0.60}_{-0.43}$\\
			\hline
			diskbb & $T_{in}\rm (keV)$ & $0.47\pm {0.01}$ & $0.47\pm {0.01}$ & $0.46\pm {0.01}$ & $0.46\pm {0.01}$& $0.46\pm {0.01}$ \\
			 & $N_{diskbb}$ & $56986.55^{+2170.17}_{-2100.60}$ & $58042.18^{+1575.57}_{-1525.28}$ & $59675.14^{+1510.41}_{-1465.52}$ & $58593.56^{+1952.74}_{-1878.95}$ & $59736.40^{+1752.87}_{-1677.46}$\\
			 & $C_{ME}$ & $1.00^{+0.15}_{-0.13}$ & $1.18^{+0.17}_{-0.15}$ & $1.19^{+0.13}_{-0.12}$ & $1.03^{+0.13}_{-0.10}$& $1.28^{+0.14}_{-0.12}$ \\
			 & $\chi^{2}/\nu$ & $950.50 /991$ & $1003.96 /991$ & $1033.12 /991$ & $1017.55 /991$ & $1039.67 /991$ \\
			 & $\chi^{2}_{\nu}$ & $0.96 $ & $1.01 $ & $1.04 $ & $1.03 $& $1.05 $ \\
			\hline
		\end{tabular}
	    \end{table*}
     
\subsection{Uncertainty analysis}\label{EA}   
In the above fittings, we used the BH mass of $M_{BH} = 11\pm 2 \ M_{\odot}$, the distance of $D = 3.39\pm 0.34\ \rm kpc$ \citep{lamer2021giant}, and the inclination of $i = 29.2^{+0.3}_{-0.5}\ \rm degrees$ \citep{jia2022detailed} as fiducial values. However, the BH mass, the distance from the source, the inclination for this source are still quite uncertain. The minimum estimated black hole mass is about 8 $M_{\odot}$ \citep{jana2020accretion}, the maximum one of 13 $M_{\odot}$ \citep{lamer2021giant}; the distance to the source could be from about 2.2 kpc \citep{chauhan2021measuring} to about 4 kpc \citep{lamer2021giant}. The inclination angle of the binary has been estimated as the range of $\sim 25^{\circ} - 46^{\circ}$ \citep{jia2022detailed,carotenuto2022black,bhowmick2022spectral,wu2022accretion}, based on the independent measurements. \cite{titarchuk2022maxi} suggested a large inclination using a correlation technique, however this marginal correlation and scaling method will induce a large uncertainty, so that in the following we still choose the inclination range between $25^{\circ}-46^{\circ}$. With the different parameter spaces, the values of X-ray luminosity and Eddington luminosity can be calculated, the accretion luminosity of the BH system is generally below $6\%$ of Eddington luminosity. 

\begin{figure}
    \centering
    \includegraphics[scale=0.5]{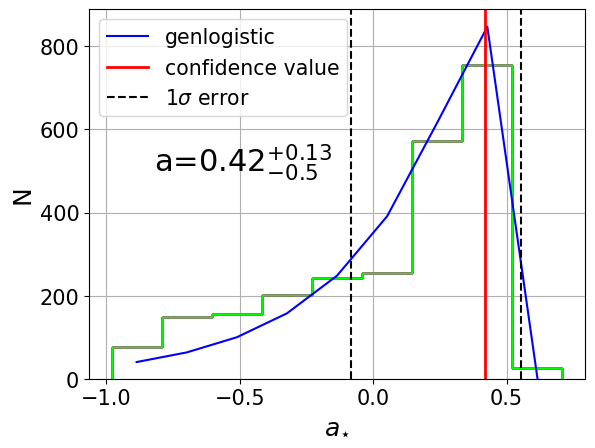}
    \caption{The distribution of $a_{\star}$ from the Monte Carlo method. The black dash lines illustrate 1 $\sigma$ error, the red line for the best-fitting value, and the blue line for the genlogistic distribution fitting.}
    \label{stat}
\end{figure}

Finally, following the similar Monte Carlo method by \cite{guan2021physical,zhao2021estimating}, we set over 2000 sets of parameters: the inclination from $25^{\circ}$ to $46^{\circ}$, the black hole mass from 8 $M_{\odot}$ to 13 $M_{\odot}$, the distance from 1.6 kpc to 4 kpc, evenly distributed to test the influence by different parameters, then, we use the model 2 to figure out the distribution of the $a_{\star}$. The final distribution of $a_{\star}$ is shown in Fig. \ref{stat}, which we use the python package fitter \citep{thomas_cokelaer_2022_7080297} to fit. The best fitting function is genlogistic (generalized logistic) distribution: $f(x,c)=c\frac{exp(-x)}{(1+exp(-x))^{c+1}}$, where c=0.05, giving $a_{\star}=0.42^{+0.13}_{-0.50}$ with 68.3\% (1 $\sigma$ error) confidence level. 

\begin{figure}
    \centering
    \includegraphics[scale=0.5]{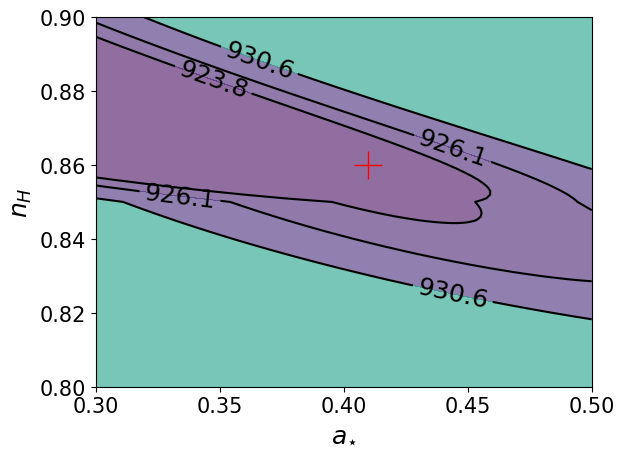}
    \caption{This contour figure is shown for iterating over $N_{\rm H}$ in the range of $(0.8- 0.9)\times 10^{21}$ cm$^{-2}$ and $a_{\star}$ from 0.3 to 0.5. The red cross represents the best fit with $\chi^{2} \sim 921.6$, the black line $\chi^{2} \sim 923.8$ for $1\sigma$, the black line $\chi^{2} \sim 926.1$ for $2\sigma$, the black line $\chi^{2} \sim 930.6$ for $3\sigma$.}
    \label{nh}
\end{figure}

In the previous spectral fittings, we have fixed the column density $N_{\rm H}$ at a normal value. We also tested the impact of this parameter, when $N_{\rm H}$ is free, on fitting the spin parameter $a_{\star}$ which has been illustrated in Fig. \ref{nh}. Although the column density $N_{\rm H}$ is hard to restrict, the best-fit results still give $a_{\star}\sim$ 0.41 and $N_{\rm H}\sim 8.6\times10^{21}\rm cm^{-2}$. The former researches \citep{wang20222018,zdziarski2022insight,li2023orbit} have shown that the difference of the absolute flux between Insight-HXMT and NuSTAR is within a few percent, which leads to the spin uncertainty $\lesssim$ 0.01, that is still within statistical uncertainty.

\section{Conclusion and Discussion}\label{conclusion}
In this paper, we reported the results of the continuum spectrum analysis about MAXI J1348-630 with an energy range 2 -20 keV based on Insight-HXMT observations during the 2019 outburst. As the fiducial value of the input parameters, the distance of the source $D = 3.39 \rm kpc$, the mass of the black hole $M = 11 M_{\odot}$ and the inclination of the system $i= 29.2^\circ$, the average spin of the black hole is determined as $0.41 \pm 0.03$ from the observation data in very soft state modeled by kerrbb, with the inner disk temperature at $0.47 \pm 0.01 \rm keV$ and the apparent inner disk radius at $5.33 \pm 0.10 R_{g}$. Furthermore, due to the quite uncertainty of the input parameters $D, M, i$, we tested the influence of these three parameters with Monte Carlo methods by setting over 2000 set parameters, e.g., $D: 1.6 - 4 \rm kpc$, $M: 8 - 13 \rm M_{\odot}$, and $i: 25^{\circ} - 46^{\circ}$ so that the best-fitted distribution showed in Fig. \ref{EA} still confirms a moderate spin black hole of $a_{\star}=0.42^{+0.13}_{-0.50}$ (1 $\sigma$ error) in MAXI J1348-630.

For the test, we use the convolution model {\em simpl} to replace the {\em powerlaw} model to make the continuum fittings. The fitting results for ObsID 403 are $a_{\star}\sim 0.41\pm 0.01$, $\dot{M}\sim 0.97\pm 0.01$, $\Gamma\sim 2.6^{+0.2}_{-1.9}$, hardening factor $\sim 0.016\pm 0.003$, which is consistent with those of Model 2. Because the difference between the convolution model {\em simpl} and additive model {\em powerlaw} would concentrate on the bands below $\sim 2$ keV \citep{steiner2009simple}, our data only covering from 2 - 20 keV will not be affected. In addition, we also tested the convolution model {\em thcomp} replacing {\em nthcomp} model in the fittings and got similar results.

\begin{figure}
    \centering
    \includegraphics[scale=0.55]{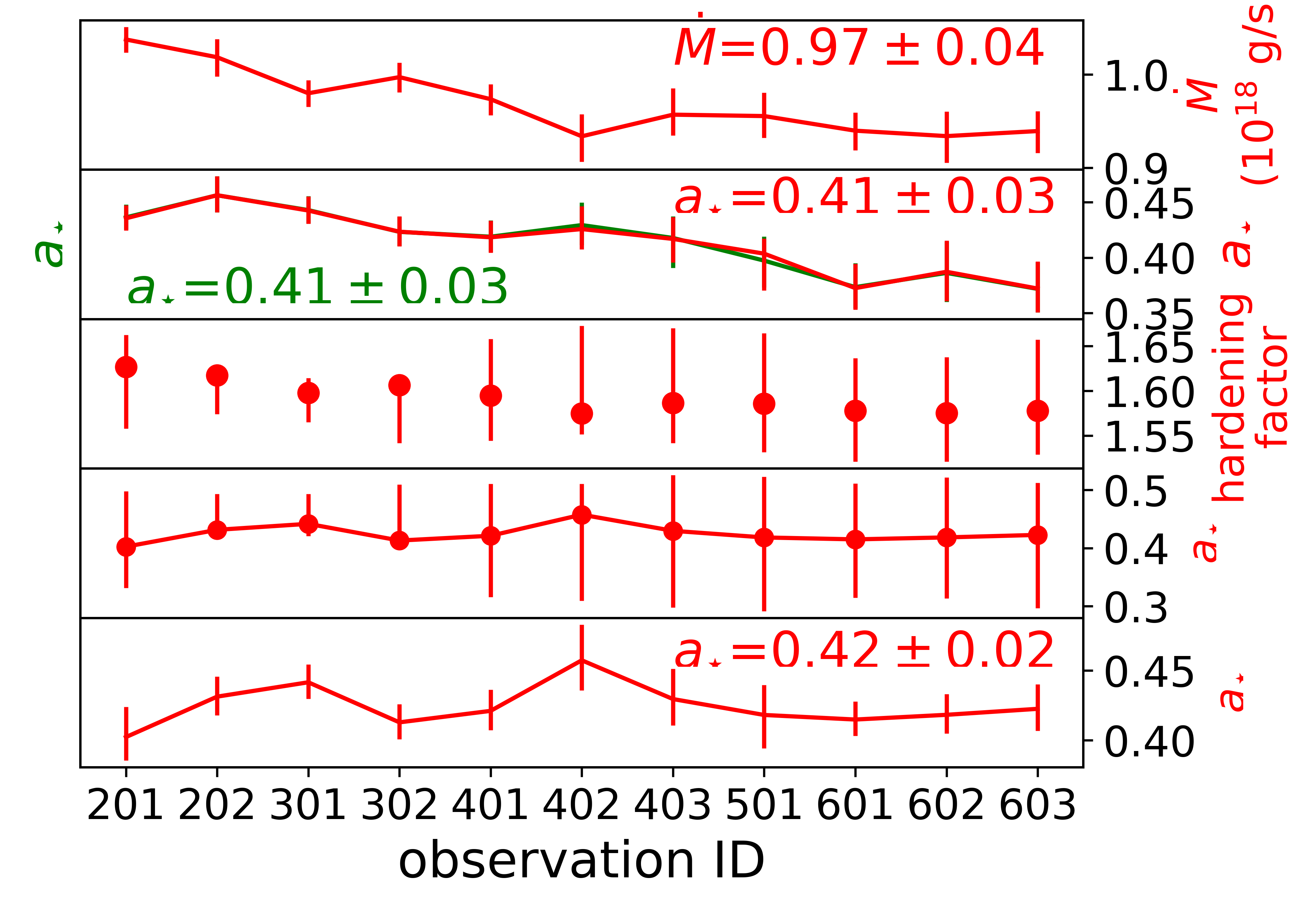}
    \caption{Comparison of the fitted parameters for two different models: 2 and 3. In the top panel: the red line is for the effective accretion rate $\dot{M}$ in units of $10^{18}\rm g/s$. In the second panel: the green line is for $a_{\star}$ in model 2, and the red line is for $a_{\star}$ in model 3. In the third panel: in the spectral fittings, we allow the hardening factor to vary with the effective accretion rate $\dot{M}$, which declines with the declining luminosity and the mean hardening factor is found to be 1.59 and the error bars are given by iterating over hardening factor and $a_{\star}$ with 1 $\sigma$. In the fourth panel: The spin parameter $a_{\star}$ is determined when the hardening factor changes as in the third panel and the error bars are given by iterating over hardening factor and $a_{\star}$ with 1 $\sigma$. Since $a_{\star}$ and the hardening factor degenerate strongly and the error bars are large in the third and fourth panels. In the bottom panel: We fix the hardening factor at the center value of the traversal results, the spin parameter $a_{\star}$ with small error bars is illustrated in this panel. The BH spin does not show any significant evolution with the accretion rate with the mean value of $a_{\star}\sim 0.42\pm 0.02$.}
    \label{compare}
\end{figure}

In the continuum spectral fittings, we have taken four models to check the influence of the disc and non-thermal models (also see Fig. \ref{compare}). From the comparison between model 2 and model 3, the influence of different non-thermal components could be ignored, and $a_{\star}=0.41\pm 0.03$ from kerrbb also agreed with the results of the $a_{\star}$ distribution $0.42^{+0.13}_{-0.50}$ in section \ref{EA}. From the fitting results of model 1 and model 4, there is also little influence on the disk component from the non-thermal components. The dikkbb model gets consistent results: the mean apparent inner disk radius $R_{in}\sim 5.33\pm 0.10\ R_{g}$, and the inner disk temperature $T_{in}\sim 0.47\pm 0.01\ \rm keV$.

We noticed that the measured spin of the black hole declines over time slightly, $a_{\star}\sim 0.45$ to $\sim 0.37$. In the former 8 observations, the change of $a_{\star}$ is not obvious, but in the last 3 observations, the spin $a_{\star}$ falls down below 0.4. In kerrbb model, the effective accretion rate $\dot{M}$ and $a_{\star}$ are strongly related (the Pearson product-moment correlation coefficient is 0.73), suggesting the trend of an apparently declining spin with declining luminosity or accretion rate. Since the spectral hardening can change with the luminosity \citep{davis2006testing}, and we have fixed the hardening factor to be 1.6 in the previous fittings, so we here let the hardening factor be a free parameter and then study the effect on the measured values of $a_{\star}$. We make the spectral fittings again with the hardening factor varying from 1.5 to 1.7 and the spin $a_{\star}$ from 0.3 to 0.5 in all the observations. In the third panel of Fig. \ref{compare}, the hardening factor change from 1.62 to 1.58 with the declining luminosity, and then in the same time, the derived $a_{\star}$ does not show the declining trend with the observations, which is illustrated in the two bottom panels of Fig. \ref{compare}.

\section*{Acknowledgements}
We are grateful to the referee for the useful comments and suggestions. This work is supported by the National Key Research and Development Program of China (Grants No. 2021YFA0718503), the NSFC (12133007, U1838103). This work has made use of data from the \textit{Insight}-HXMT mission, a project funded by China National Space Administration (CNSA) and the Chinese Academy of Sciences (CAS).

%%%%%%%%%%%%%%%%%%%%%%%%%%%%%%%%%%%%%%%%%%%%%%%%%%
\section*{Data Availability}
Data that were used in this paper are from Institute of High Energy Physics Chinese Academy of Sciences(IHEP-CAS) and are publicly available for download from the \textit{Insight}-HXMT website. To process and fit the spectrum, this research has made use of XRONOS and FTOOLS provided by NASA.



\bibliographystyle{mnras}
\bibliography{example} % if your bibtex file is called example.bib

\begin{thebibliography}{}
\makeatletter
\relax
\def\mn@urlcharsother{\let\do\@makeother \do\$\do\&\do\#\do\^\do\_\do\%\do\~}
\def\mn@doi{\begingroup\mn@urlcharsother \@ifnextchar [ {\mn@doi@}
  {\mn@doi@[]}}
\def\mn@doi@[#1]#2{\def\@tempa{#1}\ifx\@tempa\@empty \href
  {http://dx.doi.org/#2} {doi:#2}\else \href {http://dx.doi.org/#2} {#1}\fi
  \endgroup}
\def\mn@eprint#1#2{\mn@eprint@#1:#2::\@nil}
\def\mn@eprint@arXiv#1{\href {http://arxiv.org/abs/#1} {{\tt arXiv:#1}}}
\def\mn@eprint@dblp#1{\href {http://dblp.uni-trier.de/rec/bibtex/#1.xml}
  {dblp:#1}}
\def\mn@eprint@#1:#2:#3:#4\@nil{\def\@tempa {#1}\def\@tempb {#2}\def\@tempc
  {#3}\ifx \@tempc \@empty \let \@tempc \@tempb \let \@tempb \@tempa \fi \ifx
  \@tempb \@empty \def\@tempb {arXiv}\fi \@ifundefined
  {mn@eprint@\@tempb}{\@tempb:\@tempc}{\expandafter \expandafter \csname
  mn@eprint@\@tempb\endcsname \expandafter{\@tempc}}}

\bibitem[\protect\citeauthoryear{Alabarta, M{\'e}ndez, Garc{\'\i}a, Peirano,
  Altamirano, Zhang  \& Karpouzas}{Alabarta
  et~al.}{2022}]{alabarta2022variability}
Alabarta K.,  M{\'e}ndez M.,  Garc{\'\i}a F.,  Peirano V.,  Altamirano D.,
  Zhang L.,   Karpouzas K.,  2022, Monthly Notices of the Royal Astronomical
  Society

\bibitem[\protect\citeauthoryear{Bardeen, Press  \& Teukolsky}{Bardeen
  et~al.}{1972}]{bardeen1972rotating}
Bardeen J.~M.,  Press W.~H.,   Teukolsky S.~A.,  1972, The Astrophysical
  Journal, 178, 347

\bibitem[\protect\citeauthoryear{Belloni, Zhang, Kylafis, Reig  \&
  Altamirano}{Belloni et~al.}{2020}]{belloni2020time}
Belloni T.~M.,  Zhang L.,  Kylafis N.~D.,  Reig P.,   Altamirano D.,  2020,
  Monthly Notices of the Royal Astronomical Society, 496, 4366

\bibitem[\protect\citeauthoryear{Bhowmick, Debnath, Jana, Chatterjee,
  Chatterjee  \& Nath}{Bhowmick et~al.}{2022}]{bhowmick2022spectral}
Bhowmick R.,  Debnath D.,  Jana A.,  Chatterjee D.,  Chatterjee K.,   Nath
  S.~K.,  2022, 44th COSPAR Scientific Assembly. Held 16-24 July, 44, 2256

\bibitem[\protect\citeauthoryear{Blandford \& Payne}{Blandford \&
  Payne}{1982}]{blandford1982hydromagnetic}
Blandford R.,  Payne D.,  1982, Monthly Notices of the Royal Astronomical
  Society, 199, 883

\bibitem[\protect\citeauthoryear{Blandford \& Znajek}{Blandford \&
  Znajek}{1977}]{blandford1977electromagnetic}
Blandford R.~D.,  Znajek R.~L.,  1977, Monthly Notices of the Royal
  Astronomical Society, 179, 433

\bibitem[\protect\citeauthoryear{Cangemi, Rodriguez, Belloni, Gouiff{\`e}s,
  Grinberg, Laurent, Petrucci  \& Wilms}{Cangemi
  et~al.}{2022}]{cangemi2022integral}
Cangemi F.,  Rodriguez J.,  Belloni T.,  Gouiff{\`e}s C.,  Grinberg V.,
  Laurent P.,  Petrucci P.-O.,   Wilms J.,  2022, arXiv preprint
  arXiv:2210.08561

\bibitem[\protect\citeauthoryear{Cao et~al.,}{Cao et~al.}{2020}]{cao2020medium}
Cao X.,  et~al., 2020, SCIENCE CHINA Physics, Mechanics \& Astronomy, 63, 1

\bibitem[\protect\citeauthoryear{Carotenuto, Tremou, Corbel, Fender, Woudt  \&
  Miller-Jones}{Carotenuto et~al.}{2019}]{carotenuto2019meerkat}
Carotenuto F.,  Tremou E.,  Corbel S.,  Fender R.,  Woudt P.,   Miller-Jones
  J.,  2019, The Astronomer's Telegram, 12497, 1

\bibitem[\protect\citeauthoryear{Carotenuto, Tetarenko  \& Corbel}{Carotenuto
  et~al.}{2022a}]{carotenuto2022modelling}
Carotenuto F.,  Tetarenko A.,   Corbel S.,  2022a, Monthly Notices of the Royal
  Astronomical Society, 511, 4826

\bibitem[\protect\citeauthoryear{Carotenuto, Corbel  \& Tzioumis}{Carotenuto
  et~al.}{2022b}]{carotenuto2022black}
Carotenuto F.,  Corbel S.,   Tzioumis A.,  2022b, Monthly Notices of the Royal
  Astronomical Society: Letters, 517, L21

\bibitem[\protect\citeauthoryear{Chakraborty, Ratheesh, Bhattacharyya, Tomsick,
  Tombesi, Fukumura  \& Jaisawal}{Chakraborty
  et~al.}{2021}]{chakraborty2021nustar}
Chakraborty S.,  Ratheesh A.,  Bhattacharyya S.,  Tomsick J.~A.,  Tombesi F.,
  Fukumura K.,   Jaisawal G.~K.,  2021, Monthly Notices of the Royal
  Astronomical Society, 508, 475

\bibitem[\protect\citeauthoryear{Chauhan et~al.,}{Chauhan
  et~al.}{2021}]{chauhan2021measuring}
Chauhan J.,  et~al., 2021, Monthly Notices of the Royal Astronomical Society:
  Letters, 501, L60

\bibitem[\protect\citeauthoryear{Chen et~al.,}{Chen et~al.}{2020}]{chen2020low}
Chen Y.,  et~al., 2020, Science China Physics, Mechanics \& Astronomy, 63, 1

\bibitem[\protect\citeauthoryear{Cokelaer et~al.,}{Cokelaer
  et~al.}{2022}]{thomas_cokelaer_2022_7080297}
Cokelaer T.,  et~al., 2022, cokelaer/fitter: v1.5.1,
  \mn@doi{10.5281/zenodo.7080297}, \url
  {https://doi.org/10.5281/zenodo.7080297}

\bibitem[\protect\citeauthoryear{Dabrowski, Fabian, Iwasawa, Lasenby  \&
  Reynolds}{Dabrowski et~al.}{1997}]{dabrowski1997profile}
Dabrowski Y.,  Fabian A.,  Iwasawa K.,  Lasenby A.,   Reynolds C.,  1997,
  Monthly Notices of the Royal Astronomical Society, 288, L11

\bibitem[\protect\citeauthoryear{Dauser, Garc{\'\i}a, Parker, Fabian  \&
  Wilms}{Dauser et~al.}{2014}]{dauser2014role}
Dauser T.,  Garc{\'\i}a J.,  Parker M.,  Fabian A.,   Wilms J.,  2014, Monthly
  Notices of the Royal Astronomical Society: Letters, 444, L100

\bibitem[\protect\citeauthoryear{Davis, Done  \& Blaes}{Davis
  et~al.}{2006}]{davis2006testing}
Davis S.~W.,  Done C.,   Blaes O.~M.,  2006, The Astrophysical Journal, 647,
  525

\bibitem[\protect\citeauthoryear{Denisenko et~al.,}{Denisenko
  et~al.}{2019}]{denisenko2019optical}
Denisenko D.,  et~al., 2019, The Astronomer's Telegram, 12430, 1

\bibitem[\protect\citeauthoryear{Fabian, Rees, Stella  \& White}{Fabian
  et~al.}{1989}]{fabian1989x}
Fabian A.,  Rees M.,  Stella L.,   White N.~E.,  1989, Monthly Notices of the
  Royal Astronomical Society, 238, 729

\bibitem[\protect\citeauthoryear{Garc{\'\i}a, Dauser, Steiner, McClintock, Keck
   \& Wilms}{Garc{\'\i}a et~al.}{2015}]{garcia2015estimating}
Garc{\'\i}a J.~A.,  Dauser T.,  Steiner J.~F.,  McClintock J.~E.,  Keck M.~L.,
   Wilms J.,  2015, The Astrophysical Journal Letters, 808, L37

\bibitem[\protect\citeauthoryear{Guan et~al.,}{Guan
  et~al.}{2021}]{guan2021physical}
Guan J.,  et~al., 2021, Monthly Notices of the Royal Astronomical Society, 504,
  2168

\bibitem[\protect\citeauthoryear{Jana, Debnath, Chatterjee, Chakrabarti,
  Chatterjee  \& Bhowmick}{Jana et~al.}{2019}]{jana2019preliminary}
Jana A.,  Debnath D.,  Chatterjee D.,  Chakrabarti S.~K.,  Chatterjee K.,
  Bhowmick R.,  2019, The Astronomer's Telegram, 12505, 1

\bibitem[\protect\citeauthoryear{Jana, Debnath, Chatterjee, Chatterjee,
  Chakrabarti, Naik, Bhowmick  \& Kumari}{Jana
  et~al.}{2020}]{jana2020accretion}
Jana A.,  Debnath D.,  Chatterjee D.,  Chatterjee K.,  Chakrabarti S.~K.,  Naik
  S.,  Bhowmick R.,   Kumari N.,  2020, The Astrophysical Journal, 897, 3

\bibitem[\protect\citeauthoryear{Jia et~al.,}{Jia
  et~al.}{2022}]{jia2022detailed}
Jia N.,  et~al., 2022, Monthly Notices of the Royal Astronomical Society, 511,
  3125

\bibitem[\protect\citeauthoryear{Jithesh, Misra, Maqbool  \& Mall}{Jithesh
  et~al.}{2021}]{jithesh2021broad}
Jithesh V.,  Misra R.,  Maqbool B.,   Mall G.,  2021, Monthly Notices of the
  Royal Astronomical Society, 505, 713

\bibitem[\protect\citeauthoryear{Kennea \& Negoro}{Kennea \&
  Negoro}{2019}]{kennea2019maxi}
Kennea J.,  Negoro H.,  2019, The Astronomer's Telegram, 12434, 1

\bibitem[\protect\citeauthoryear{Kerr}{Kerr}{1963}]{kerr1963gravitational}
Kerr R.~P.,  1963, Physical review letters, 11, 237

\bibitem[\protect\citeauthoryear{Kubota, Tanaka, Makishima, Ueda, Dotani, Inoue
   \& Yamaoka}{Kubota et~al.}{1998}]{kubota1998evidence}
Kubota A.,  Tanaka Y.,  Makishima K.,  Ueda Y.,  Dotani T.,  Inoue H.,
  Yamaoka K.,  1998, Publications of the Astronomical Society of Japan, 50, 667

\bibitem[\protect\citeauthoryear{Kulkarni et~al.,}{Kulkarni
  et~al.}{2011}]{kulkarni2011measuring}
Kulkarni A.~K.,  et~al., 2011, Monthly Notices of the Royal Astronomical
  Society, 414, 1183

\bibitem[\protect\citeauthoryear{Kumar, Bhattacharyya, Bhatt  \& Misra}{Kumar
  et~al.}{2022}]{kumar2022estimation}
Kumar R.,  Bhattacharyya S.,  Bhatt N.,   Misra R.,  2022, Monthly Notices of
  the Royal Astronomical Society, 513, 4869

\bibitem[\protect\citeauthoryear{Lamer, Schwope, Predehl, Traulsen, Wilms  \&
  Freyberg}{Lamer et~al.}{2021}]{lamer2021giant}
Lamer G.,  Schwope A.,  Predehl P.,  Traulsen I.,  Wilms J.,   Freyberg M.,
  2021, Astronomy \& Astrophysics, 647, A7

\bibitem[\protect\citeauthoryear{Li, Zimmerman, Narayan  \& McClintock}{Li
  et~al.}{2005}]{li2005multitemperature}
Li L.-X.,  Zimmerman E.~R.,  Narayan R.,   McClintock J.~E.,  2005, The
  Astrophysical Journal Supplement Series, 157, 335

\bibitem[\protect\citeauthoryear{Li et~al.,}{Li et~al.}{2023}]{li2023orbit}
Li X.,  et~al., 2023, arXiv preprint arXiv:2302.10714

\bibitem[\protect\citeauthoryear{Liao et~al.,}{Liao
  et~al.}{2020}]{liao2020background}
Liao J.-Y.,  et~al., 2020, Journal of High Energy Astrophysics, 27, 24

\bibitem[\protect\citeauthoryear{Liu et~al.,}{Liu et~al.}{2020}]{liu2020high}
Liu C.,  et~al., 2020, SCIENCE CHINA Physics, Mechanics \& Astronomy, 63, 1

\bibitem[\protect\citeauthoryear{Liu et~al.,}{Liu
  et~al.}{2022}]{liu2022transitions}
Liu H.,  et~al., 2022, The Astrophysical Journal, 938, 108

\bibitem[\protect\citeauthoryear{Makishima, Maejima, Mitsuda, Bradt, Remillard,
  Tuohy, Hoshi  \& Nakagawa}{Makishima
  et~al.}{1986}]{makishima1986simultaneous}
Makishima K.,  Maejima Y.,  Mitsuda K.,  Bradt H.,  Remillard R.,  Tuohy I.,
  Hoshi R.,   Nakagawa M.,  1986, The Astrophysical Journal, 308, 635

\bibitem[\protect\citeauthoryear{Mall, Vadakkumthani  \& Misra}{Mall
  et~al.}{2022}]{mall2022broadband}
Mall G.,  Vadakkumthani J.,   Misra R.,  2022, Research in Astronomy and
  Astrophysics

\bibitem[\protect\citeauthoryear{Matsuoka et~al.,}{Matsuoka
  et~al.}{2009}]{matsuoka2009maxi}
Matsuoka M.,  et~al., 2009, Publications of the Astronomical Society of Japan,
  61, 999

\bibitem[\protect\citeauthoryear{McClintock, Shafee, Narayan, Remillard, Davis
  \& Li}{McClintock et~al.}{2006}]{mcclintock2006spin}
McClintock J.~E.,  Shafee R.,  Narayan R.,  Remillard R.~A.,  Davis S.~W.,   Li
  L.-X.,  2006, The Astrophysical Journal, 652, 518

\bibitem[\protect\citeauthoryear{McClintock, Narayan  \& Steiner}{McClintock
  et~al.}{2013}]{mcclintock2013black}
McClintock J.~E.,  Narayan R.,   Steiner J.~F.,  2013, in , The Physics of
  Accretion onto Black Holes.
Springer, pp 295--322

\bibitem[\protect\citeauthoryear{Misner, Thorne  \& Wheeler}{Misner
  et~al.}{1973}]{misner1973gravitation}
Misner C.~W.,  Thorne K.,   Wheeler J.,  1973, San Francisco, p.~660

\bibitem[\protect\citeauthoryear{Mitsuda et~al.,}{Mitsuda
  et~al.}{1984}]{mitsuda1984energy}
Mitsuda K.,  et~al., 1984, Publications of the Astronomical Society of Japan,
  36, 741

\bibitem[\protect\citeauthoryear{Nied{\'z}wiecki, Szanecki  \&
  Zdziarski}{Nied{\'z}wiecki et~al.}{2019}]{niedzwiecki2019improved}
Nied{\'z}wiecki A.,  Szanecki M.,   Zdziarski A.~A.,  2019, Monthly Notices of
  the Royal Astronomical Society, 485, 2942

\bibitem[\protect\citeauthoryear{{Novikov} \& {Thorne}}{{Novikov} \&
  {Thorne}}{1973}]{novikov1973astrophysics}
{Novikov} I.~D.,  {Thorne} K.~S.,  1973, Black Holes (Les Astres Occlus), pp
  343--450

\bibitem[\protect\citeauthoryear{Panizo-Espinar et~al.,}{Panizo-Espinar
  et~al.}{2022}]{panizo2022discovery}
Panizo-Espinar G.,  et~al., 2022, arXiv preprint arXiv:2205.09128

\bibitem[\protect\citeauthoryear{Penna, McKinney, Narayan, Tchekhovskoy, Shafee
   \& McClintock}{Penna et~al.}{2010}]{penna2010simulations}
Penna R.~F.,  McKinney J.~C.,  Narayan R.,  Tchekhovskoy A.,  Shafee R.,
  McClintock J.~E.,  2010, Monthly Notices of the Royal Astronomical Society,
  408, 752

\bibitem[\protect\citeauthoryear{Remillard \& McClintock}{Remillard \&
  McClintock}{2006}]{remillard2006x}
Remillard R.~A.,  McClintock J.~E.,  2006, Annu. Rev. Astron. Astrophys., 44,
  49

\bibitem[\protect\citeauthoryear{Reynolds \& Nowak}{Reynolds \&
  Nowak}{2003}]{reynolds2003fluorescent}
Reynolds C.~S.,  Nowak M.~A.,  2003, Physics Reports, 377, 389

\bibitem[\protect\citeauthoryear{Russell, Anderson, Miller-Jones, Degenaar,
  Eijnden, Sivakoff  \& Tetarenko}{Russell et~al.}{2019}]{russell2019atca}
Russell T.,  Anderson G.,  Miller-Jones J.,  Degenaar N.,  Eijnden J.~v.,
  Sivakoff G.~R.,   Tetarenko A.,  2019, The astronomer's telegram, 12456, 1

\bibitem[\protect\citeauthoryear{Saha, Pal, Mandal  \& Manna}{Saha
  et~al.}{2021}]{saha2021multi}
Saha D.,  Pal S.,  Mandal M.,   Manna A.,  2021, arXiv preprint
  arXiv:2104.09926

\bibitem[\protect\citeauthoryear{Sanna et~al.,}{Sanna
  et~al.}{2019}]{sanna2019nicer}
Sanna A.,  et~al., 2019, The astronomer's telegram, 12447, 1

\bibitem[\protect\citeauthoryear{Shafee, McClintock, Narayan, Davis, Li  \&
  Remillard}{Shafee et~al.}{2005}]{shafee2005estimating}
Shafee R.,  McClintock J.~E.,  Narayan R.,  Davis S.~W.,  Li L.-X.,   Remillard
  R.~A.,  2005, The Astrophysical Journal, 636, L113

\bibitem[\protect\citeauthoryear{Shafee, McKinney, Narayan, Tchekhovskoy,
  Gammie  \& McClintock}{Shafee et~al.}{2008}]{shafee2008three}
Shafee R.,  McKinney J.~C.,  Narayan R.,  Tchekhovskoy A.,  Gammie C.~F.,
  McClintock J.~E.,  2008, The Astrophysical Journal, 687, L25

\bibitem[\protect\citeauthoryear{Shimura \& Takahara}{Shimura \&
  Takahara}{1995}]{shimura1995spectral}
Shimura T.,  Takahara F.,  1995, The Astrophysical Journal, 445, 780

\bibitem[\protect\citeauthoryear{Steiner, Narayan, McClintock  \&
  Ebisawa}{Steiner et~al.}{2009}]{steiner2009simple}
Steiner J.~F.,  Narayan R.,  McClintock J.~E.,   Ebisawa K.,  2009,
  Publications of the Astronomical Society of the Pacific, 121, 1279

\bibitem[\protect\citeauthoryear{Steiner, McClintock, Remillard, Gou, Yamada
  \& Narayan}{Steiner et~al.}{2010}]{steiner2010constant}
Steiner J.~F.,  McClintock J.~E.,  Remillard R.~A.,  Gou L.,  Yamada S.,
  Narayan R.,  2010, The Astrophysical Journal Letters, 718, L117

\bibitem[\protect\citeauthoryear{Steiner et~al.,}{Steiner
  et~al.}{2011}]{steiner2011spin}
Steiner J.~F.,  et~al., 2011, Monthly Notices of the Royal Astronomical
  Society, 416, 941

\bibitem[\protect\citeauthoryear{Titarchuk \& Seifina}{Titarchuk \&
  Seifina}{2022}]{titarchuk2022maxi}
Titarchuk L.,  Seifina E.,  2022, arXiv preprint arXiv:2211.06271

\bibitem[\protect\citeauthoryear{Tominaga et~al.,}{Tominaga
  et~al.}{2020}]{tominaga2020discovery}
Tominaga M.,  et~al., 2020, The Astrophysical Journal Letters, 899, L20

\bibitem[\protect\citeauthoryear{Verner, Ferland, Korista  \& Yakovlev}{Verner
  et~al.}{1996}]{verner1996atomic}
Verner D.,  Ferland G.~J.,  Korista K.,   Yakovlev D.,  1996, arXiv preprint
  astro-ph/9601009

\bibitem[\protect\citeauthoryear{Wang et~al.,}{Wang
  et~al.}{2022}]{wang20222018}
Wang P.,  et~al., 2022, Monthly Notices of the Royal Astronomical Society, 512,
  4541

\bibitem[\protect\citeauthoryear{Weng, Cai, Zhang, Zhang, Chen, Huang  \&
  Tao}{Weng et~al.}{2021}]{weng2021time}
Weng S.-S.,  Cai Z.-Y.,  Zhang S.-N.,  Zhang W.,  Chen Y.-P.,  Huang Y.,   Tao
  L.,  2021, The Astrophysical Journal Letters, 915, L15

\bibitem[\protect\citeauthoryear{Wilms, Allen  \& McCray}{Wilms
  et~al.}{2000}]{wilms2000absorption}
Wilms J.,  Allen A.,   McCray R.,  2000, The Astrophysical Journal, 542, 914

\bibitem[\protect\citeauthoryear{Wu, Wang  \& Sai}{Wu
  et~al.}{2023}]{wu2022accretion}
Wu H.,  Wang W.,   Sai N.,  2023, Journal of High Energy Astrophysics, 37, 25

\bibitem[\protect\citeauthoryear{Yatabe et~al.,}{Yatabe
  et~al.}{2019}]{yatabe2019maxi}
Yatabe F.,  et~al., 2019, The Astronomer's Telegram, 12425, 1

\bibitem[\protect\citeauthoryear{Zdziarski, Johnson  \& Magdziarz}{Zdziarski
  et~al.}{1996}]{zdziarski1996broad}
Zdziarski A.~A.,  Johnson W.~N.,   Magdziarz P.,  1996, Monthly Notices of the
  Royal Astronomical Society, 283, 193

\bibitem[\protect\citeauthoryear{Zdziarski, Szanecki, Poutanen, Gierli{\'n}ski
  \& Biernacki}{Zdziarski et~al.}{2020}]{zdziarski2020spectral}
Zdziarski A.~A.,  Szanecki M.,  Poutanen J.,  Gierli{\'n}ski M.,   Biernacki
  P.,  2020, Monthly Notices of the Royal Astronomical Society, 492, 5234

\bibitem[\protect\citeauthoryear{Zdziarski, You, Szanecki, Li  \& Ge}{Zdziarski
  et~al.}{2022}]{zdziarski2022insight}
Zdziarski A.~A.,  You B.,  Szanecki M.,  Li X.-B.,   Ge M.,  2022, The
  Astrophysical Journal, 928, 11

\bibitem[\protect\citeauthoryear{Zhang, Cui  \& Chen}{Zhang
  et~al.}{1997}]{zhang1997black}
Zhang S.~N.,  Cui W.,   Chen W.,  1997, The Astrophysical Journal, 482, L155

\bibitem[\protect\citeauthoryear{Zhang et~al.,}{Zhang
  et~al.}{2020a}]{zhang2020overview}
Zhang S.-N.,  et~al., 2020a, Science China Physics, Mechanics \& Astronomy, 63,
  1

\bibitem[\protect\citeauthoryear{Zhang et~al.,}{Zhang
  et~al.}{2020b}]{zhang2020nicer}
Zhang L.,  et~al., 2020b, Monthly Notices of the Royal Astronomical Society,
  499, 851

\bibitem[\protect\citeauthoryear{Zhang et~al.,}{Zhang
  et~al.}{2022}]{zhang2022peculiar}
Zhang W.,  et~al., 2022, The Astrophysical Journal, 927, 210

\bibitem[\protect\citeauthoryear{Zhao et~al.,}{Zhao
  et~al.}{2020}]{zhao2020confirming}
Zhao X.-S.,  et~al., 2020, Journal of High Energy Astrophysics, 27, 53

\bibitem[\protect\citeauthoryear{Zhao et~al.,}{Zhao
  et~al.}{2021}]{zhao2021estimating}
Zhao X.,  et~al., 2021, The Astrophysical Journal, 916, 108

\makeatother
\end{thebibliography}

%\begin{thebibliography}{99}
%\end{thebibliography}

%%%%%%%%%%%%%%%%%%%%%%%%%%%%%%%%%%%%%%%%%%%%%%%%%%

%%%%%%%%%%%%%%%%% APPENDICES %%%%%%%%%%%%%%%%%%%%%

%\appendix


%If you want to present additional material which would interrupt the flow of the main paper, it can be placed in an Appendix which appears after the list of references.

%%%%%%%%%%%%%%%%%%%%%%%%%%%%%%%%%%%%%%%%%%%%%%%%%%

% Don't change these lines
\bsp	% typesetting comment
\label{lastpage}
\end{document}